%% file: main.tex
\documentclass[runningheads]{llncs}

\usepackage[mobile]{eccv}

\usepackage{eccvabbrv}

\input{settings/packages}
\input{settings/newcommands}

\usepackage[accsupp]{axessibility}  

\usepackage{hyperref}

\usepackage{orcidlink}

\begin{document}
\include{acronyms}

\newlength{\captionvertspace}
\setlength{\captionvertspace}{0.0cm}

\newlength{\captionpostvertspace}
\setlength{\captionpostvertspace}{0.0cm}

\title{Concise Plane Arrangements for Low-Poly\\Surface and Volume Modelling} 



\author{Raphael Sulzer \and
Florent Lafarge}

\authorrunning{R.~Sulzer \& F.~Lafarge}

\institute{Centre Inria d'Universit\'e C\^ote d'Azur\\
\email{firstname.lastname@inria.fr}
}

\maketitle

\begin{abstract}
  Plane arrangements are a useful tool for surface and volume modelling. However, their main drawback is poor scalability. We introduce two key novelties that enable the construction of plane arrangements for complex objects and entire scenes: (i) an ordering scheme for the plane insertion and (ii) the direct use of input points during arrangement construction. Both ingredients reduce the number of unwanted splits, resulting in improved scalability of the construction mechanism by up to two orders of magnitude compared to existing algorithms. We further introduce a remeshing and simplification technique that allows us to extract low-polygon surface meshes and lightweight convex decompositions of volumes from the arrangement.
  We show that our approach leads to state-of-the-art results for the aforementioned tasks
  by comparing it to learning-based and traditional approaches on various different datasets. Our implementation is available at \url{https://github.com/raphaelsulzer/compod}.
  \keywords{plane arrangement \and low-poly \and reconstruction}
\end{abstract}

\input{sections/1_intro}
\input{sections/2_related-work}
\input{sections/3_algorithms}
\input{sections/4_experiments}

\input{sections/5_conclusion}

%
%
\bibliographystyle{splncs04}
\bibliography{references}
\end{document}

%% file: settings/packages.tex
\usepackage{multirow}
\usepackage{makecell}
\usepackage{booktabs,array}
\usepackage{balance}
\usepackage{wrapfig}
\usepackage{alphalph}

\usepackage{pgfplots}
\pgfplotsset{compat=1.17}
\usepackage{tikz}
\usetikzlibrary{calc,spy,patterns,arrows.meta,positioning,shapes,shapes.misc,backgrounds}
\usepackage{wrapfig}
\usepackage{enumitem}
\usepackage{adjustbox}
\usepackage{colortbl}

\usepackage{placeins}
\usepackage{acronym}
\usepackage{graphicx}
\usepackage{amsmath}
\usepackage{amssymb}
\usepackage{booktabs}
\usepackage{algorithm}
\usepackage[noend]{algpseudocode}
\usepackage{wrapfig}
\usepackage{multirow}
\usepackage[permil]{overpic}

%% file: settings/newcommands.tex
\renewcommand*{\eqref}[1]{Eq.~\hyperref[#1]{\ref*{#1}}}
\newcommand*{\chref}[1]{Chapter~\hyperref[#1]{\ref*{#1}}}
\newcommand*{\secref}[1]{Sec.~\hyperref[#1]{\ref*{#1}}}
\newcommand*{\sectionref}[1]{Section~\hyperref[#1]{\ref*{#1}}}
\newcommand*{\figref}[1]{Fig.~\hyperref[#1]{\ref*{#1}}}
\newcommand*{\figureref}[1]{Figure~\hyperref[#1]{\ref*{#1}}}
\newcommand*{\tabref}[1]{Tab.~\hyperref[#1]{\ref*{#1}}}
\newcommand*{\tableref}[1]{Table~\hyperref[#1]{\ref*{#1}}}
\newcommand*{\Subref}[1]{\hyperref[#1]{(\subref*{#1})}}
\renewcommand*{\algref}[1]{Alg.~\hyperref[#1]{\ref*{#1}}}
\newcommand*{\algorithmref}[1]{Algorithm~\hyperref[#1]{\ref*{#1}}}
\newcolumntype{H}{>{\setbox0=\hbox\bgroup}c<{\egroup}@{}}

\newcommand{\chamfer}{CD}
\newcommand{\chamfersq}{$\text{CD}^2$}
\newcommand{\hausdorff}{HD}
\newcommand{\normal}{NC}

\newcommand{\graph}{G}
\newcommand{\tree}{T}

\DeclareMathOperator*{\argmax}{arg\,max}

\newcommand{\RAPH}[1]{\textcolor{black}{#1}}

\makeatletter
\newcommand{\definetrim}[2]{%
  \define@key{Gin}{#1}[]{\setkeys{Gin}{trim=#2,clip}}%
}
\makeatother

\makeatletter
\newcommand{\definescale}[2]{%
  \define@key{Gin}{#1}[]{\setkeys{Gin}{scale=#2}}%
}
\makeatother

\makeatletter
\newcommand{\gridon}[2]{%
  \define@key{Gin}{#1}[]{\setkeys{Gin}{#2}}%
}
\makeatother

\algdef{SE}[DOWHILE]{Do}{doWhile}{\algorithmicdo}[1]{\algorithmicwhile\ #1}%

%% file: acronyms.tex

\newacro{2d}[$2$D]{two-dimensional}
\newacro{3d}[$3$D]{three-dimensional}
\newacro{2dt}[$2$DT]{2D Delaunay triangulation}
\newacro{3dt}[$3$DT]{3D Delaunay tetrahedralisation}
\newacro{dt}[DT]{Delaunay triangulation}

\newacro{aabb}[AABB]{axis alligned bounding box}

\newacro{pc}[PC]{polyhedral complex}

\newacro{bsp}[BSP]{binary space partitioning}
\newacro{csg}[CSG]{constructive solid geometry}
\newacro{cad}[CAD]{computer-aided design}
\newacro{ksr}[KSR]{Kinetic Shape Reconstruction}

\newacro{psr}[PSR]{Poisson Surface Reconstruction}
\newacro{spsr}[SPSR]{Screened Poisson Surface Reconstruction}

\newacro{ier}[IER]{intrinsic-extrinsic ratio}
\newacro{igr}[IGR]{Implicit Geometric Regularisation}
\newacro{lig}[LIG]{Local Implicit Grids}
\newacro{p2m}[P2M]{Point2Mesh}
\newacro{sap}[SAP]{Shape As Points}
\newacro{p2s}[P2S]{Points2Surf}
\newacro{onet}[ONet]{Occupancy Networks}
\newacro{conet}[ConvONet]{Convolutional Occupancy Networks}
\newacro{dgnn}[DGNN]{Delaunay-Graph Neural Network}
\newacro{poco}[POCO]{Point Convolution for Surface Reconstruction}
\newacro{rlpm}[RLPM]{Robust Low-Poly Meshing}

\newacro{iou}[IoU]{intersection over union}
\newacro{cd}[CD]{Chamfer distance}
\newacro{nc}[NC]{normal consistency}
\newacro{noc}[NoC]{number of components}

\newacro{pca}[PCA]{principal component analysis}
\newacro{mlp}[MLP]{multilayer perceptron}

\newacro{bce}[BCE]{binary cross entropy}
\newacro{mse}[MSE]{mean square error}

\newacro{cnn}[CNN]{convolutional neural network}
\newacroplural{cnn}[CNNs]{convolutional neural networks}

\newacro{gnn}[GNN]{graph neural network}
\newacroplural{gnn}[GNNs]{graph neural networks}

\newacro{lod}[LOD]{level of detail}
\newacroplural{lod}[LODs]{levels of detail}
\newacro{mvs}[MVS]{multi-view stereo}
\newacro{sfm}[SfM]{structure from motion}
\newacro{lidar}[LiDAR]{Light Detection and Ranging}
\newacro{als}[ALS]{airborne laser scanning}

\newacro{nerf}[NeRF]{neural radiance field}

\newacro{sdf}[SDF]{signed distance function}
\newacro{of}[OF]{occupancy function}

\newacro{fcn}[FCN]{fully connected network}
\newacro{dsr}[DSR]{deep surface reconstruction}

\newacro{mise}[MISE]{multi-resolution isosurface extraction}

\newacro{tft}[TFT]{triangle-from-tetrahedra}


%% file: sections/1_intro.tex
\section{Introduction}
\label{sec-intro}

\input{figures/teaser/teaser_figure}

Explicit 3D mesh representations such as tetrahedralisations or triangle surface meshes allow for a good approximation of freeform geometry using atomic parts of surface and volume. However, \RAPH{for storing, analysing or manipulating 3D data, these dense representations} introduce a computational burden by describing even simple shapes with many redundant elements. 
Lightweight polygon surface meshes composed of few polygons or volume meshes composed of few convex polyhedra can facilitate the handling of the underlying 3D data.
However, the reconstruction of such low-poly meshes from raw data measurements is still a challenging task.
To this end, we present a scalable plane arrangement method for low-poly surface and volume modelling from point clouds. 

Plane arrangement methods \cite{bauchet2020ksr,ChauveCVPR2010,nan_iccv2017} use planar shapes detected from data measurements and arrange them to form a polyhedral decomposition, \ie a partition of a 3D domain by polyhedra. Subsequently, a concise mesh can be extracted from the partition by selecting a subset of the polygonal facets or polyhedral cells.
This strategy is particularly interesting as (i) it can be used for both surface and volume modelling, (ii) it offers a good robustness to imperfect data and (iii) it comes with desirable geometric guarantees such as \RAPH{watertightness and orientability of the surface and convexity of the polyhedral cells.
The main shortcoming of plane arrangement methods is their poor scalability.} The \emph{exhaustive} plane arrangement iteratively slices the 3D domain with detected planes which leads to a time complexity of
$\mathcal{O}(n^3)$, with $n$ being the number of input planes \cite{Edelsbrunner83}.
The \emph{adaptive} arrangement uses convex planar polygons~\cite{Murali:1997:CSA} or axis-aligned bounding boxes~\cite{chen2022points2poly} instead of infinite planes, which drastically reduces the number of intersection computations and the complexity of the resulting arrangement.
However, in practice, none of the existing mechanisms can process more than \RAPH{1000} planar primitives without parallelisation schemes.

The key contribution of our work is an efficient mechanism for constructing more concise arrangements in less time than existing methods. We directly exploit input points to avoid unnecessary splitting operations and replace intersection tests between polyhedral cells and polygons with simply point-plane orientation tests.
\RAPH{We then carefully order the plane insertion operations to further lower the computational complexity of the algorithm.}
\RAPH{We also introduce a remeshing and simplification strategy that reduces the number of extracted polygonal surface facets or the number of extracted convex polyhedra from the arrangement.}
\RAPH{The combination of these ingredients allows us to produce concise representations of complex objects and entire scenes from several thousand input planar shapes (cf. \figref{fig:teaser}).
We empirically demonstrate the effectiveness of our algorithm by comparing to previous plane arrangement, mesh simplification and mesh decomposition methods on various datasets.
}

%% file: figures/teaser/teaser_figure.tex
\begin{figure}
\centering
\includegraphics[width=\textwidth,trim=0 0 0 0, clip]{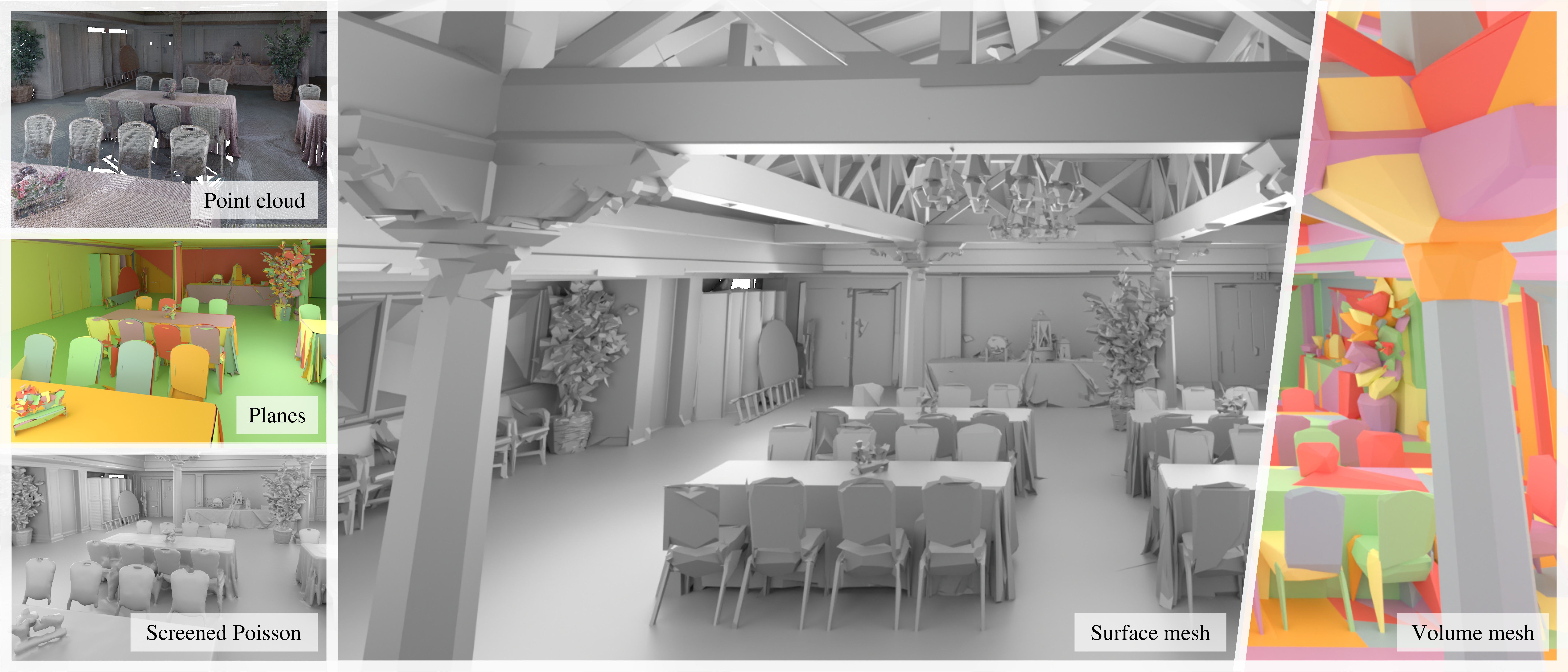}
\vspace{\captionvertspace}
\caption{Our representation of the Meeting Room~\cite{tanksandtemples} as a watertight and intersection-free surface mesh with 90k polygonal facets (center) or as a simplified volume mesh with 2500 convex polyhedra (right). Our method inputs 40k planes (middle left) detected from a LiDAR scan of the scene (top left). For comparison, a Screened Poisson \cite{screened_poisson} reconstruction consists of over 6M triangles (bottom left).}
\vspace{\captionpostvertspace}
\label{fig:teaser}
\end{figure}

%% file: sections/2_related-work.tex
\section{Related Work}

Our review of related work covers algorithms for concise surface reconstruction, convex decomposition of 3D shapes and construction of plane arrangements.

\subsubsection{Concise Surface Reconstruction.} One common way for reconstructing a concise surface mesh from data measurements consists in generating a dense triangle mesh from point clouds or multiview stereo images before reducing its number of triangle facets. While the first step relies upon a vast literature \cite{sulzer2024survey} and recent advances on NeRF, mesh simplification is mainly based on a few geometry processing algorithms that iteratively collapse edges \cite{Garland_97,salinas_cgf15,chen2023robust} or group triangle facets into planar clusters before remeshing \cite{vsa}. 
These algorithms are fast and scalable, but deliver triangle meshes that usually fail to preserve the structure of objects. In contrast, plane assembly methods detect planar shapes from point clouds \cite{survey_boubekeur} and assemble them to form a polygonal surface mesh. Assembling can be done by considering the dual of an adjacency graph between planar shapes. However, the construction of such a graph is unlikely to be consistent and requires to be interactively completed \cite{thijs_2011,arikan12}. More robust assembly approaches rather decompose the 3D space into polyhedral cells with splitting operations induced by the planar shapes. The polygonal surface mesh is then extracted by a binary labelling of cells \cite{bauchet2020ksr,ChauveCVPR2010,mura_cgf16} or facets \cite{boulch2013piecewise,fang_cvpr20,nan_iccv2017} from the decomposition. The construction of such plane arrangements has a high algorithmic complexity and no existing mechanism can process more than 1000 planar shapes without block decomposition schemes that introduce border artefacts. Our work follows an arrangement approach, but with an efficient construction mechanism which is able to handle up to two orders of magnitude more planar shapes.


\subsubsection{Plane Arrangement.} The trivial way to construct a plane arrangement is to iteratively slice the 3D domain with the supporting plane of each detected planar shape \cite{Edelsbrunner83}. Such an \emph{exhaustive} plane arrangement mechanism typically produces dense plane arrangements as even supporting planes of spatially distant polygons are likely to intersect. The exhaustive construction becomes intractable with only a few hundred shapes \cite{nan_iccv2017}. 
To improve on complexity and performance, the slicing operations can be restricted to the polyhedral cells that include the associated planar shape only \cite{Murali:1997:CSA}. This condition can be relaxed by initially dilating the polygon enclosing the shape \cite{ChauveCVPR2010} or by testing inclusion from the bounding box of cells \cite{chen2022points2poly}. In such an \emph{adaptive} plane arrangement, the ordering in which input shapes are processed impacts both the quality of the output decomposition and the performance of its construction. However, ordering schemes proposed in the literature remain simple and application-driven, \eg large \cite{Murali:1997:CSA} or vertical \cite{chen2022points2poly} shapes first. 
The \emph{kinetic} mechanism KSR \cite{bauchet2020ksr} does not perform slicing operations on polyhedral cells in an iterative manner. Instead, it grows 2D polygons at a constant speed until they collide and form polyhedral cells. 
KSR produces a concise decomposition without proximity rules, but the collisions are costly to simulate.
Our method follows an adaptive plane arrangement construction but with an efficient ordering scheme and the exploitation of raw measurements to significantly gain in performance and compactness.

\subsubsection{Volumetric Decomposition of 3D shapes.} Exact convex decompositions of 3D shapes typically produce a prohibitively large number of convex cells (in short, convexes) \cite{chazelle1984convex,bajaj1992convex,hershberger1998erased}, while
approximate convex decompositions offer a trade-off between the fidelity to the input shape and the number of convexes from the decomposition \cite{wei2022coacd,mamou2016vhacd}. Traditionally, axis-aligned split planes are used to partition the input shape into convex, or nearly convex parts \cite{wei2022coacd,mamou2016vhacd}. The decompositions are refined and simplified by moving cutting planes and grouping cells \cite{wei2022coacd}.
In our approach, we use cutting planes detected on the object surface instead, to achieve a high fidelity to the input shape. Furthermore, we efficiently group cells of our plane arrangement to produce a decomposition with low complexity.

Some recent neural network based methods also produce decompositions from other types of primitives such as cubes and cylinders \cite{yu2023dualcsg,ren2021stump} or permit convexes of the decomposition to overlap \cite{chen2020bspnet,deng2020cvxnet}.  However, non-convex and overlapping decompositions are not desirable for applications such as collision detection or point localisation \cite{wei2022coacd}. Moreover, most neural network based methods can only process small objects with simple geometries. In contrast, our method produces convex decompositions from complex objects and scenes.

%% file: sections/3_algorithms.tex
\section{Algorithm}
\label{sec_algo}

\input{figures/pipeline/pipeline_figure_eccv}

Our approach takes a point cloud representing a 3D object or scene as input (\figref{subfig:pipeline_a}) and returns either a concise, watertight and intersection-free polygon surface mesh (\figref{subfig:pipeline_d}) or a concise volume mesh composed of convex polyhedra (\figref{subfig:pipeline_e}).
First, we detect planar primitives defined as the association of supporting planes $P$ and a subset of input points, called inliers, to which supporting planes are fitted (\figref{subfig:pipeline_b}). We can use any standard plane detection method for this step \cite{cgal_shape_detection, yu2022planes}.
We then construct a concise polyhedral decomposition from the planes and their corresponding inlier points (\figref{subfig:pipeline_c}). Finally, we extract the outer boundaries of the observed object by labelling each polyhedral cell of the decomposition as inside or outside the surface.
For this step, we can use a proxy surface, \eg from a Screened Poisson \cite{screened_poisson} reconstruction of the input point cloud, a normal-based approach~\cite{bauchet2020ksr} or a deep occupancy field \cite{chen2022points2poly}.
We then apply a remeshing and simplification strategy to group facets (\figref{subfig:pipeline_f}) and cells (\figref{subfig:pipeline_g}) into larger components. 
In the following, we explain the adaptive plane arrangement, on which our algorithm is based, and present our key contributions. 

\subsection{Background on Adaptive Plane Arrangement}


The adaptive construction mechanism first introduced by Murali~\&~Funkhouser \cite{Murali:1997:CSA} converts a set of unoriented and disconnected polygons into a polyhedral decomposition, composed of a set of polyhedral cells $C$ and a set of polygonal facets~$F$.
The input polygons are typically computed as the 2D convex hulls of inlier points projected onto their associated supporting planes.
The algorithm starts by initializing the polyhedral decomposition to the 3D domain, \ie to a padded bounding box of the input polygons. Then, it performs a series of splitting operations on the polyhedral
decomposition, in which the polyhedral cells containing a polygon, or a part of it, are cut by the supporting plane of the polygon. The order in which polygons are processed is determined a priori and stored in a priority queue. In the inset below, the commonly used order, \ie larger polygons
\setlength{\columnsep}{0.04in}
\begin{wrapfigure}[5]{r}{0.38\linewidth}
    \vspace{-0.35in}
    \centering
    \begin{overpic}[width=\linewidth]{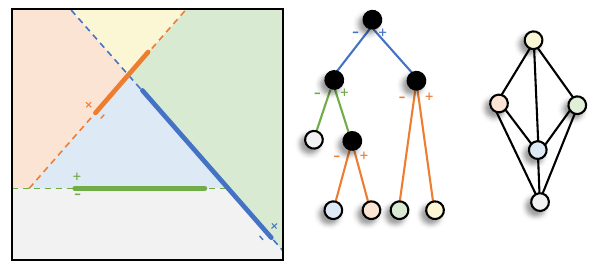}
    \put(500,20){\fontsize{6}{6}\selectfont{BSP-tree $\tree$}}
    \put(810,20){\fontsize{6}{6}\selectfont{graph $\graph$}}
    \end{overpic}
    \end{wrapfigure} %
 first, split the domain by the blue-then-green-then-orange polygon (represented by the line segments), leading to a decomposition with five polyhedral cells at the end.
A tree structure, typically a \ac{bsp}-tree, is used to track the splitting operations during the process. In such a tree (denoted by $\tree$), the root node $\tree_0$ corresponds to the 3D domain. The path from the root to a tree node~$\tree_i$ stores the splitting operations required to form the polyhedral cell $c_i$. At the end of the process, the leaves of the tree correspond to the polyhedral cells of the output decomposition. An undirected graph $\graph$, updated in tandem to the tree, is also used to store the adjacency between polyhedral cells.

\subsection{Our Concise and Scalable Plane Arrangement}
\label{sec:our_algo}

\subsubsection{Ordering of Splitting Operations.}
\label{paragraph:sorting}
In the various versions of the adaptive plane arrangement \cite{Murali:1997:CSA,ChauveCVPR2010,chen2022points2poly}, the priority queue is computed at initialization using considerations on the area or orientation of the input polygons. Unfortunately, such an ordering scheme is too simple to produce polyhedral partitions in a stable and efficient manner. We tackle this problem with a dynamic ordering scheme%
\setlength{\columnsep}{0.04in}%
\begin{wrapfigure}[5]{r}{0.38\linewidth}
    \vspace{-0.9cm}
    \centering
    \begin{overpic}[width=\linewidth]{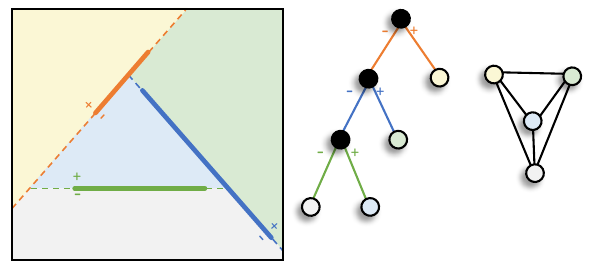}
    \put(480,20){\fontsize{6}{6}\selectfont{ BSP-tree $\tree$}}
    \put(790,20){\fontsize{6}{6}\selectfont{ graph $\graph$}}
    \end{overpic}
\end{wrapfigure}%
that seeks a decomposition with a low number of polyhedra, or equivalently a \ac{bsp}-tree with a low number of leaves. Instead of a priority queue that sorts the input polygons, we use a function that selects the next splitting operation, \ie a pair of one plane and one polyhedral cell, to be processed given the current state of the partition. It relies upon two key ideas:
\begin{enumerate}[label=\roman*.]
    \item a splitting operation that creates a polyhedral cell that cannot be split anymore, must take priority,
    \item splitting operations that create two cells containing each an as high and identical as possible number of polygons must be favored over other splits.
\end{enumerate}
The intuition behind (i) is that turning nodes into leaves in the \ac{bsp}-tree as soon as possible reduces the growth of the tree, whereas (ii) aims to reduce the number of intersection computations and balances the number of polygons in the cells of the decomposition to reduce the algorithmic complexity over the 3D domain. In the inset, continuing with the same polygon layout as in the previous example, we first split the domain along the orange polygon, \ie the only one respecting condition (i) at initialisation, then by the blue, then green polygon, also validating this condition. The BSP-tree is simpler and the decomposition has only four cells at the end.
\vspace{-0.2cm}
\subsubsection{Fast Intersection Tests with Inlier Points.}
Frequent intersection tests between polyhedra and planar primitives are necessary to identify the cells to split. These tests constitute a computational bottleneck in the various versions of the mechanism, even with approximation by bounding boxes \cite{chen2022points2poly}. We address this issue by replacing intersection tests by a simple distribution of the inlier points in the \ac{bsp}-tree. The idea is that the assignment of inlier points to the nodes of the%
\setlength{\columnsep}{0.04in}
\begin{wrapfigure}[9]{r}{0.37\linewidth}
\vspace{-0.35in}
\centering
\begin{overpic}[width=\linewidth]{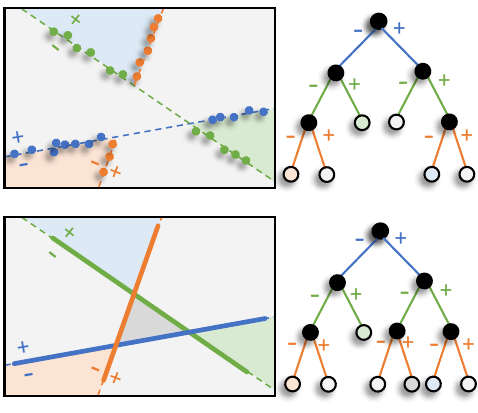}
\end{overpic}
\end{wrapfigure} %
 \ac{bsp}-tree will easily provide their relative positioning to the cutting planes. We rely on a hierarchical clustering of the inlier points based on simple point-plane orientation tests for reassigning the inlier points after a splitting operation.
This strategy leads to a strong reduction of the computational burden and directly exploits the precision of input points. This is particularly beneficial when planar shapes are concave or contain holes, as illustrated in the top decomposition of the inset. 
An arrangement strategy using only planes \cite{Edelsbrunner83} or convex polygons \cite{Murali:1997:CSA,chen2022points2poly} cannot avoid the creation of the meaningless center cell in the bottom decomposition.
\vspace{-0.2cm}
\subsubsection{Practical Description.}
We start by initialising the root node of a BSP-tree with a dilated bounding box of all inlier points. 
The root node is then associated with all inlier points and their corresponding supporting planes.
We now iterate over the leaves of the tree and check if all inlier points associated with the current node can be located on the same side of a supporting plane inside the cell.
If we find such a plane, we split the cell along it.
Otherwise, for each supporting plane $p$ we compute the sets of inlier groups $L_p$ and $R_p$ that lie fully left and right of $p$.
We then select the split that maximizes the product
$|L_p| |R_p|$, \ie we cut the current cell with the supporting plane that splits the inlier sets as equally as possible, while intersecting as few other polygons as possible. Finally, we reassign the inlier points and their associated supporting planes 
to the two new child nodes.
When there are no more polyhedra containing points, \ie no points associated with the leaves of the tree, the splitting operations stop.

\subsection{Remeshing and Simplification}
\vspace{-0.1cm}
\subsubsection{Aggregation of Convex Facets.} For surface modelling, we extract a watertight and intersection-free mesh with 
convex polygonal facets from the decomposition (\figref{subfig:pipeline_d}). 
We then collect clusters of coplanar facets and extract the boundary edges, \ie the edges that only occur once per cluster. For each cluster, we find all cycles in the set of boundary edges \cite{paton1969algorithm}. In case of multiple cycles for which one cycle is contained in another, \ie a facet with holes, we apply a 2D Delaunay triangulation constrained by the boundary edges (\figref{subfig:pipeline_f}).

\subsubsection{Aggregation of Convex Cells.}
For volume modelling, the convex cells of the reconstructed mesh can be merged into fewer, larger convex cells. In the inset, %
\begin{wrapfigure}[5]{r}{0.17\linewidth}
\vspace{-0.35in}
\centering
\includegraphics[width=\linewidth]{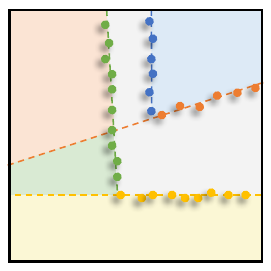}
\end{wrapfigure}%
the orange and green cell can be merged, to produce a more concise decomposition with the same geometry.
While our ordering strategy aims to reduce such unwanted splits, they cannot fully be avoided due to multiple interactions of planes for complex geometries.
We can exploit the properties of the BSP-tree in which each parent node corresponds to the union of its two children.
Starting from the leave nodes, we recursively replace two siblings by their parent cell if they lie on the same side of the surface. 
This allows us to reduce the number of cells in the decomposition by around 25\% without any geometrical computations.
Once all same-sided siblings are processed, the decomposition can still include pairs of same-sided adjacent cells $a,b$ whose union is equivalent to the convex hull of their union $\text{conv}(a \cup b)$.
We recursively test this condition for all pairs of adjacent cells of the mesh, given by the adjacency graph $G$, by comparing the sum of the volumes of the two cells with the volume of the convex hull of their union \cite{mamou2016vhacd}.
We replace adjacent cells $a$ and $b$ with their convex hull $\text{conv}(a \cup b)$, if
\vspace{-0.1cm}
\begin{equation}
\label{eq:simplification}
\lvert V_a+V_b - V_{\text{conv}(a \cup b)} \rvert < \tau,
\vspace{-0.1cm}
\end{equation}
where $V_x$ is the volume of cell $x$ and $\tau$ a positive parameter. $\tau$ determines how much we allow the volume of the new cell $\text{conv}(a \cup b)$ to differ from the combined volumes of $a$ and $b$. If $\tau = 0$ the geometry stays unaltered and the decomposition is guaranteed to be intersection free (\figref{subfig:pipeline_e}).
Relaxing $\tau$ allows to alter the geometry and the cells in the decomposition to potentially overlap (\figref{subfig:pipeline_g}). In the inset, we could then simplify further by replacing the orange, green \emph{and} yellow cell with the convex hull of their union.

%% file: figures/pipeline/pipeline_figure_eccv.tex
\begin{figure*}[t]
    \captionsetup[figure]{position=auto}
    \newcommand{\mywidth}{0.19\linewidth}
    \newcommand{\mywidthd}{0.38\linewidth}
    \definetrim{mytrim}{90 120 110 120}
    \definetrim{mytrimd}{20 70 40 70}
    
    \begin{tabular}{@{}c@{}c@{}c@{}c@{}c@{}}
    \centering
    \begin{subfigure}{\mywidth}
    \includegraphics[width=\linewidth,mytrim]{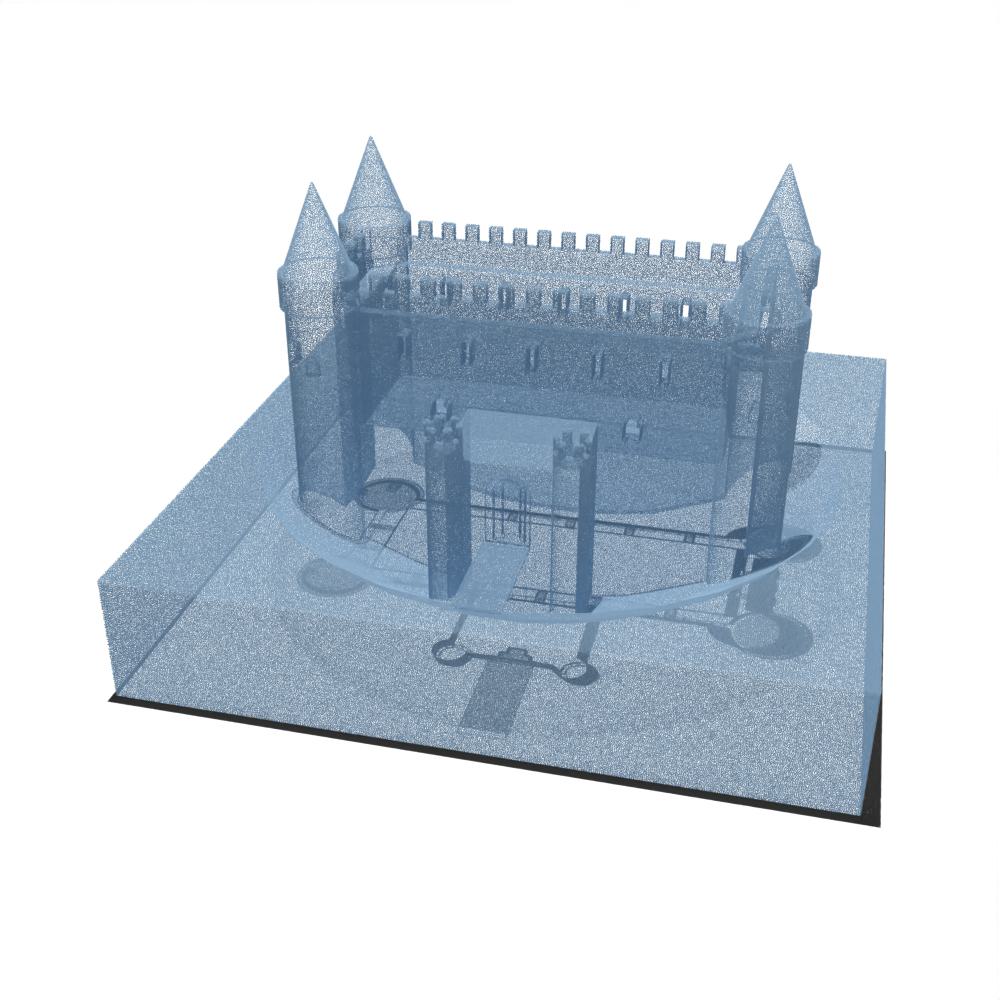}
    \vspace{-1.5\baselineskip}
    \caption{Point cloud}
    \label{subfig:pipeline_a}
    \end{subfigure}
    &
    \multicolumn{2}{c}{
    \multirow{2}{*}[5em]{
    \begin{subfigure}{\mywidthd}
    \addtocounter{subfigure}{1}
    \includegraphics[width=\linewidth,mytrimd]{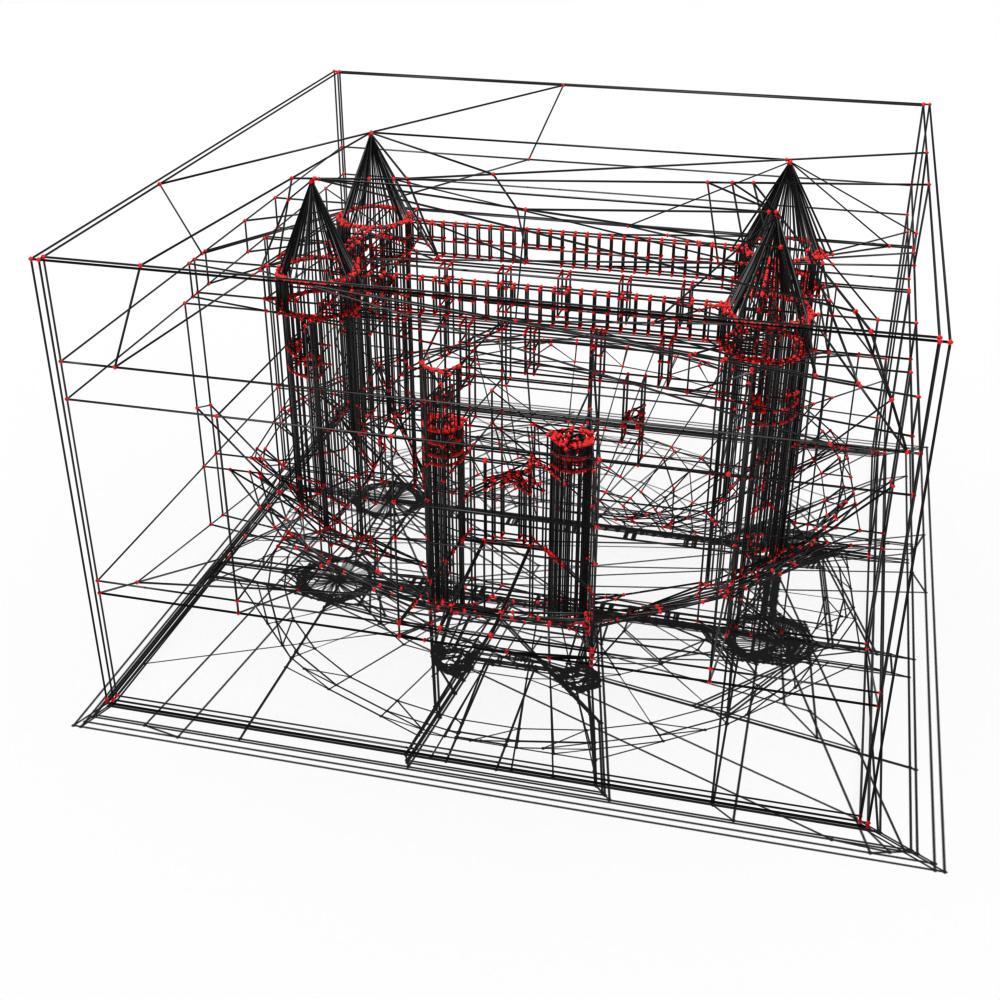}
    \vspace{-1.5\baselineskip}
    \caption{Compact plane arrangement}
    \label{subfig:pipeline_c}
    \end{subfigure}
    }}
    &
    \begin{subfigure}{\mywidth}
    \includegraphics[width=\linewidth,mytrim]{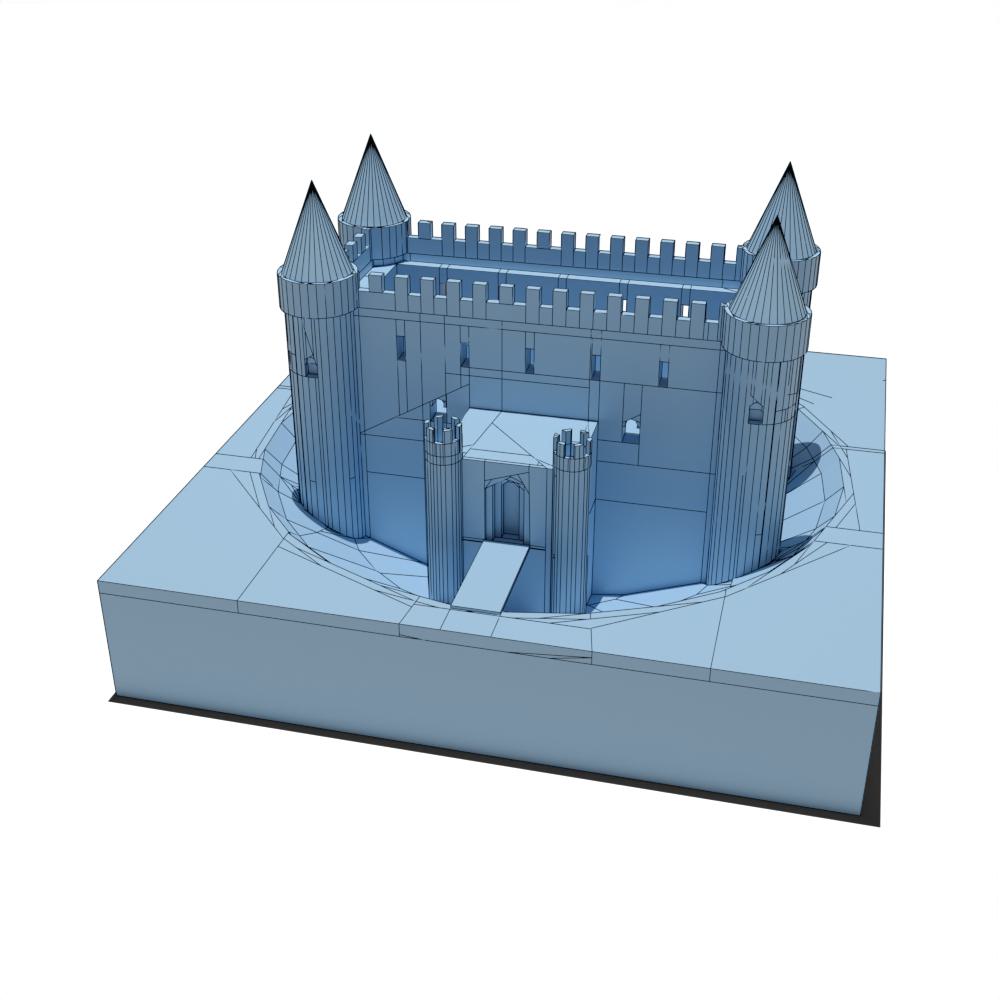}
    \vspace{-1.5\baselineskip}
    \caption{Surface mesh}
    \label{subfig:pipeline_d}
    \end{subfigure}
    &
    \begin{subfigure}{\mywidth}
    \addtocounter{subfigure}{1}
    \includegraphics[width=\linewidth,mytrim]{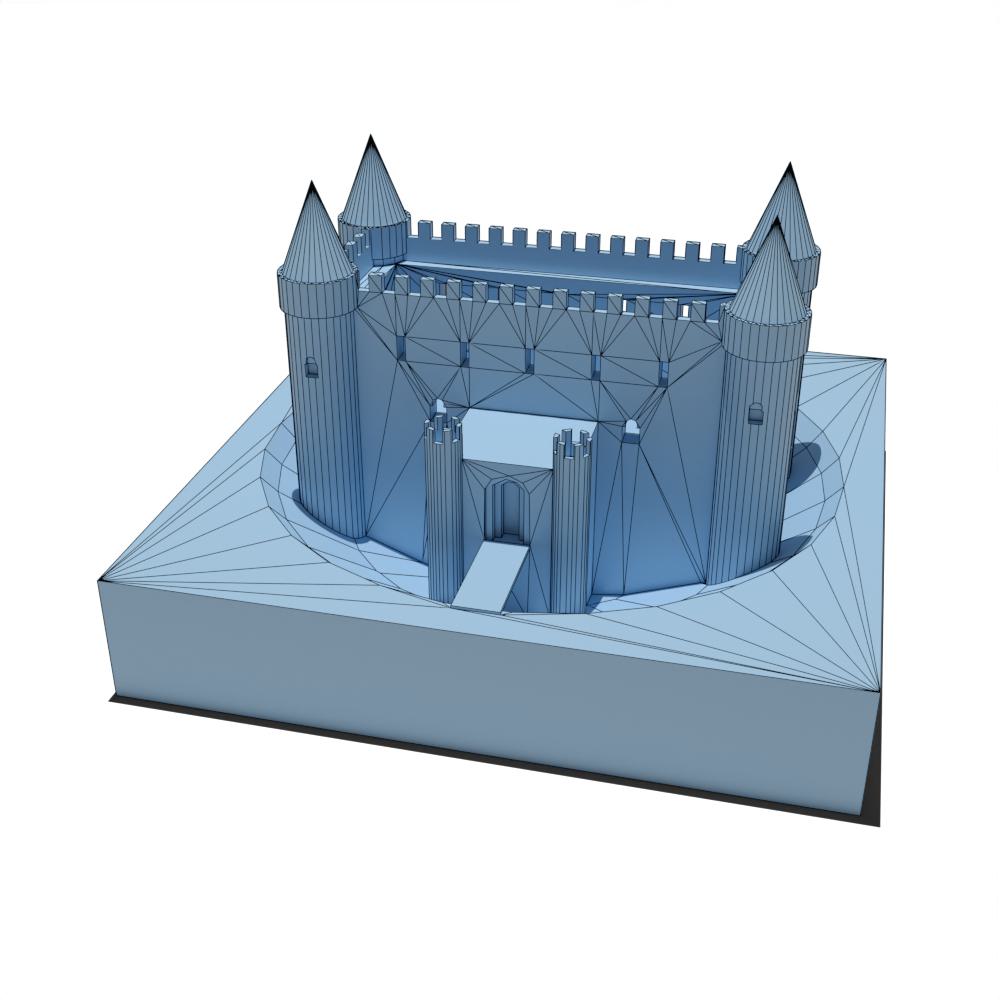}
    \vspace{-1.5\baselineskip}
    \caption{\emph{remeshed}}
    \label{subfig:pipeline_f}
    \end{subfigure}
    \\
    \begin{subfigure}{\mywidth}
    \addtocounter{subfigure}{-5}
    \includegraphics[width=\linewidth,mytrim]{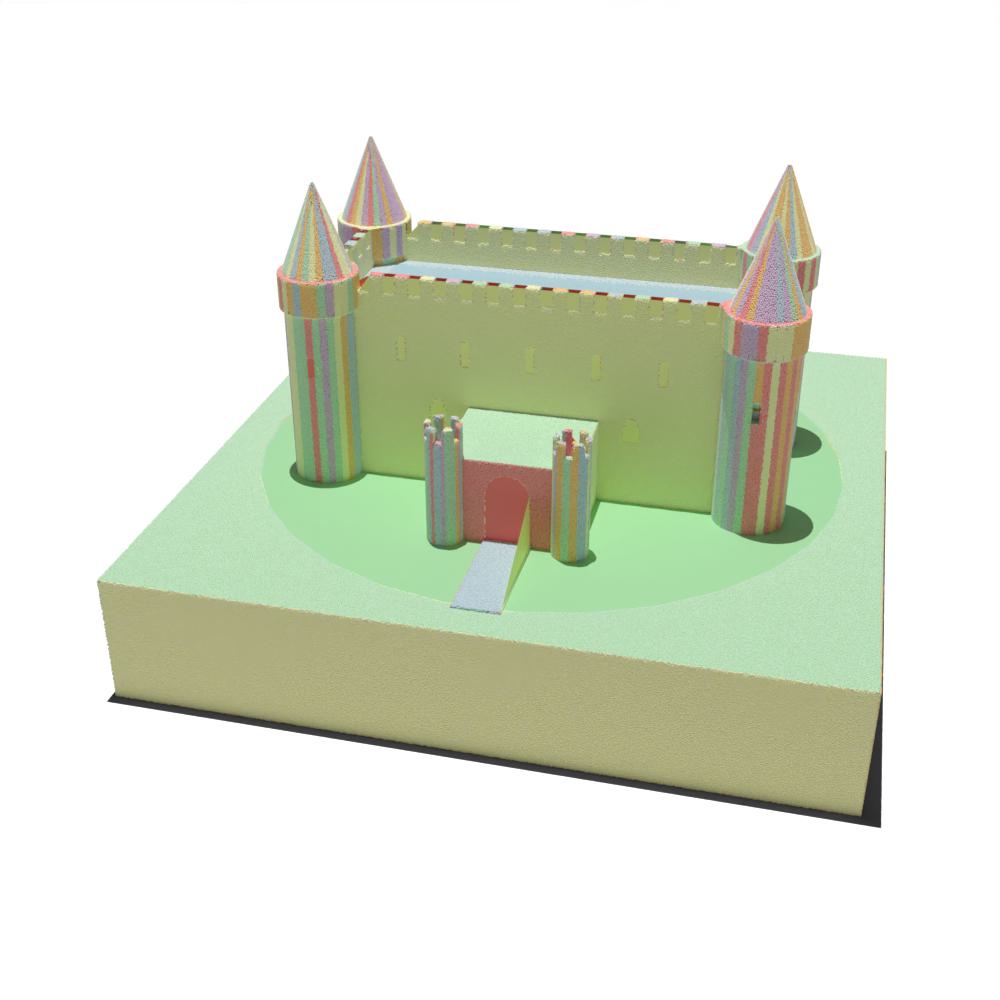}
    \vspace{-1.5\baselineskip}
    \caption{Planes}
    \label{subfig:pipeline_b}
    \end{subfigure}
    &
    &
    &
    \begin{subfigure}{\mywidth}
    \addtocounter{subfigure}{2}
    \includegraphics[width=\linewidth,mytrim]{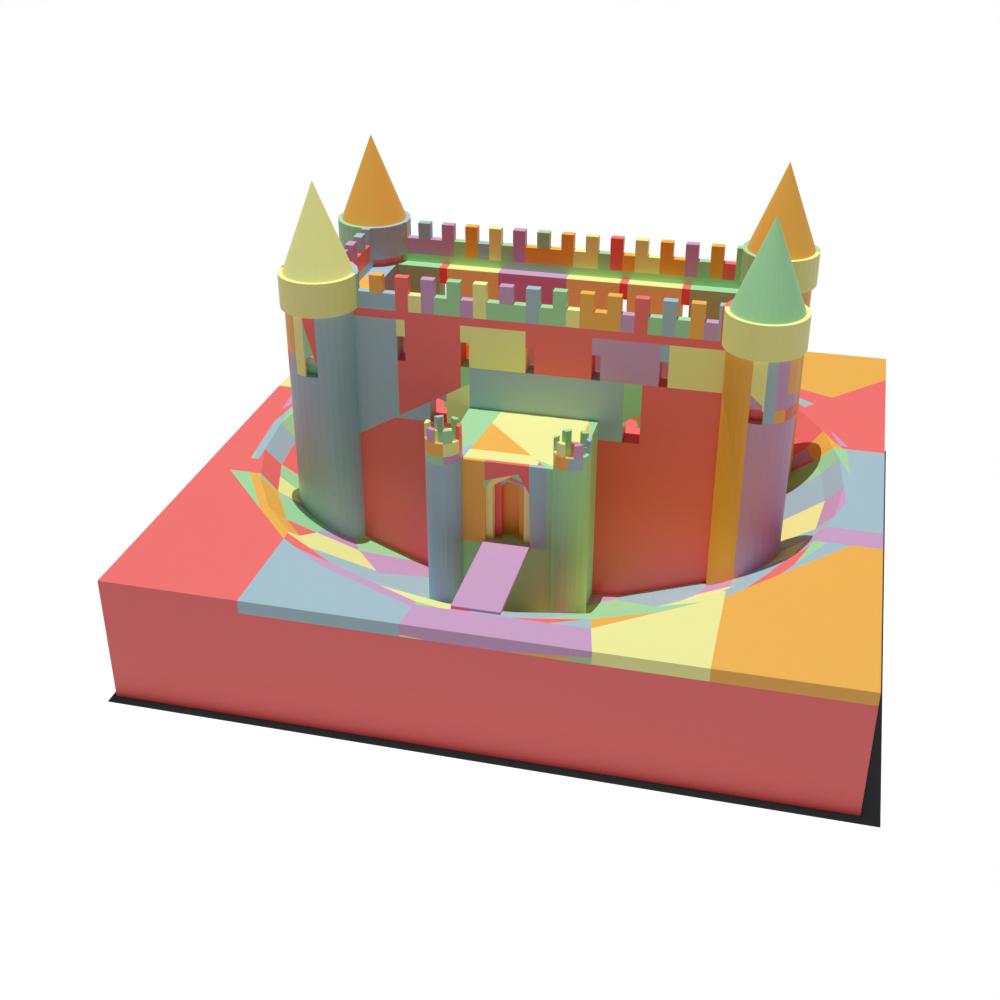}
    \vspace{-1.5\baselineskip}
    \caption{Volume mesh}
    \label{subfig:pipeline_e}
    \end{subfigure}
    &
    \begin{subfigure}{\mywidth}
    \addtocounter{subfigure}{1}
    \includegraphics[width=\linewidth,mytrim]{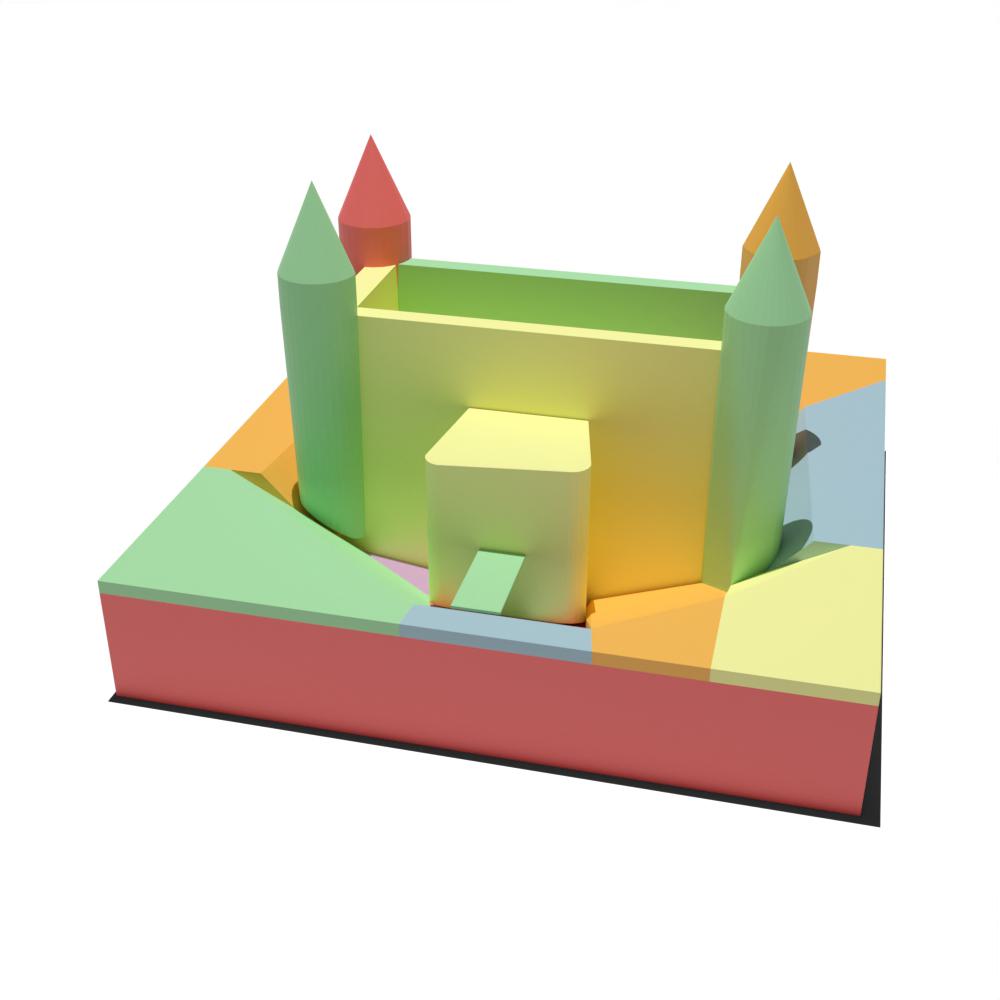}
    \vspace{-1.5\baselineskip}
    \caption{\emph{simplified}}
    \label{subfig:pipeline_g}
    \end{subfigure}

    \end{tabular}
    \vspace{\captionvertspace}
    \caption{\textbf{Pipeline.}
    Our method processes a set of planes with their corresponding inlier points \Subref{subfig:pipeline_b} detected from a point cloud \Subref{subfig:pipeline_a} into a compact plane arrangement \Subref{subfig:pipeline_c}. From this arrangement we extract either a watertight and intersection-free surface mesh \Subref{subfig:pipeline_d} or a volume mesh with intersection-free convexes \Subref{subfig:pipeline_e}.
    The surface mesh can optionally be remeshed to represent each planar region with only one facet,
    and with Delaunay triangles for regions with holes \Subref{subfig:pipeline_f}. The remeshed surface is still watertight and intersection-free.
    The volume mesh can be simplified by merging groups of convex volumes and potentially allowing them to intersect \Subref{subfig:pipeline_g}.}
    \vspace{\captionpostvertspace}
    \label{fig:pipeline}
    \end{figure*}

%% file: sections/4_experiments.tex
\section{Experiments}

In this section, we first compare the efficacy and efficiency of our plane arrangement against existing construction mechanisms and conduct an ablation study. We then evaluate the competitiveness of our general pipeline on surface and volume modelling tasks. We compare to the state-of-the-art for each task and use the 
experimental protocol
provided by the authors of the corresponding methods.

\subsection{Evaluation on Plane Arrangement Construction}

\subsubsection{Experimental Setup.}
We compare our plane arrangement against several baselines on the Thingi10k~\cite{thingi10k} dataset.
\paragraph*{Baselines.} (i) We compare to an \emph{exhaustive} arrangement which we implement by intersecting all detected input planes. (ii) We compare to an \emph{adaptive} arrangement, which we implement following ideas of the building reconstruction pipeline P2P~\cite{chen2022points2poly}. See the supplementary material for a detailed comparison of our implementation to the one of P2P. (iii) We compare to a \emph{kinetic} plane arrangement (KSR) for which we use the implementation provided by the authors. We construct arrangements with the four different mechanisms, occupancy-label the cells of the arrangement with the normal-based method of KSR and extract polygonal surface meshes from the interface of adjacet inside/outside cells.

\paragraph*{Dataset.} We use 1000 non-degenerate models randomly chosen from Thingi10k. We sample 200k points on each model and detect planar shapes \cite{yu2022planes,cgal_shape_detection} using a fitting tolerance parameter fixed to $0.8\%$ of the models' bounding box diagonal. We input the same plane configurations to all four different construction mechanisms. Because the construction mechanisms do not scale equally, we split the dataset into three groups: \emph{simple}, for models approximated by less than 100 planar shapes, \emph{moderate}, if between 100 and 250, and \emph{complex} for models with more than 250 detected planes.

\paragraph*{Metrics.} We evaluate (i) the complexity of the arrangement with the number of polyhedral cells $|C|$ and the complexity of the surface with the number of polygonal surface facets $|F_S|$,
(ii) the accuracy of the extracted surface with the symmetric Chamfer (CD) and Hausdorff (HD) distances between ground truth and reconstructed surface and
(iii) the performance of mechanisms with the construction time and memory peak.

\input{tables/thingi}

\subsubsection{Comparison.} \tableref{tab:partition} presents the quantitative results. For simple models, our algorithm offers the best complexity with more than two times less cells compared to the second best and one order magnitude less on complex ones. \figref{fig:thingi} illustrates this complexity gap on a simple model. The significant gain in complexity is not done at the expense of the quality of the reconstructed meshes as both the Hausdorff and Chamfer distances remain competitive with a lower number of polygonal facets. The exhaustive method, which performs on simple models only, offers the best accuracy, but produces overly complex meshes. KSR and Adaptive exhibit a similar accuracy to our method, but with a higher number of polygonal facets. Our method is also faster, especially on complex models. Both performance and complexity gains originate from the combination of our ordering scheme, the use of the inlier points that accurately describe the input data, and our remeshing strategy. 

\input{figures/thingi/thingi_figure}

\subsubsection{Scalability.}
We further evaluated the scalability of plane arrangements by generating configurations of planar shapes at six levels of complexity, ranging from 40 to 10k planar shapes, from six different real-world scans.
Graphs presented in \figref{fig:runtime} compare processing time and peak memory as a function of the number of input planar shapes for the different construction mechanisms. Our algorithm offers the best performances and the best stability at the different levels of complexity. The gap is particularly large for configurations with a high number of planes. Only the kinetic algorithm exhibits a lower memory peak for configurations with no more than 250 planes. 

\input{figures/runtime/runtime_figure}

\subsubsection{Ablation Study.} \tableref{tab:ablation} shows the impact of our design choices on the various evaluation metrics. 
We first examine the impact of our ordering scheme. The basic area-based sorting scheme leads to a much more complex decomposition, a less accurate surface and almost $2\times$ longer runtime.
Removing the priority condition (i) leads to a decomposition with more cells. The cells of the decomposition have less facets on average, which leads to a slightly faster runtime. However, condition (i) also seeks to better recover the general shape of an object by first inserting all planes that lie on the convex hull. Consequently, removing condition (i) leads to a worse Hausdorff distance. 
\tableref{tab:ablation} also shows the benefits of using inliers points for the intersection tests. We can reduce the complexity and construction time of the decomposition, but strongly degrade the quality of the reconstructed mesh by using only the vertices of 2D convex hulls of inlier points projected onto their supporting plane. A relevant alternative is to use 1M points uniformly sampled on the convex polygons, but this produces a more complex decomposition than our vanilla implementation.
Finally, the surface remeshing only impacts the complexity of the reconstructed mesh. Removing the cell simplification (with $\tau = 0$) impacts the complexity of the decomposition and leads to a higher runtime due to a longer surface extraction. 

\input{tables/ablation}

\subsection{Surface Mesh Simplification}

\subsubsection{Experimental Setup.}%
\input{figures/robust/combined_eccv/combined}
Besides the reconstruction of low-poly meshes directly from point clouds, another relevant application of our pipeline is the simplification of dense, high-poly surfaces into low-poly ones.
We compare our method to \acf{rlpm}~\cite{chen2023robust}.
RLMP simplifies dense surface meshes by computing an offset surface of the input and iteratively simplifying this offset.
The method establishes itself as state-of-the-art in low-poly mesh generation by comparing to over ten other low-poly meshing algorithms. We compare our method to RLMP on the dataset provided by the authors, \ie a subset of 100 models from Thingi10k \cite{thingi10k}. We exclude 45 models that have a non-orientable surface, because orientability is a requirement for the inside/outside labelling in the surface extraction step of our pipeline.  
We sample 2M points per model and use the same plane detection parameters as in the previous experiment.
Because RLPM produces triangle meshes and our method produces
polygon meshes we triangulate our output and perform edge collapse based on QEM~\cite{Garland_97} to produce models with the same number of triangles as the ones of RLPM. We call this variant \emph{OursTri}.
Note that, the authors of RLPM also experiment with replacing parts of their pipeline with QEM in their paper, which leads to worse results.

\input{figures/coacd/coacd_combined_eccv}

\subsubsection{Results.} \figref{fig:robust} shows quantitative and qualitative results of the output meshes. Our method produces polygon meshes with a similar number of polygons (Ours) and triangles (OursTri) than RLMP, but with a much better accuracy while also exhibiting a shorter runtime.
The remeshing and decimation operations of RLPM progressively degrade the accuracy on small details. In contrast, planar shapes that capture such details allow a more precise, yet concise approximation of the local geometry and facilitate the preservation of details.

\vspace{-0.3cm}
\subsection{Volume Decomposition with Intersection-Free Convexes}

\subsubsection{Experimental Setup.} Another relevant application of our pipeline is the decomposition of volumes into a low number of intersection-free convexes. We compare our pipeline with the specialized method CoACD \cite{wei2022coacd} on 50 challenging objects and scenes from \emph{Thingiverse}. We use the implementation provided by the authors. CoACD cuts an input solid mesh with equally spaced axis-alligned planes to produce a polyhedral cell decomposition. The cells are then merged into larger cells using a multi-step tree search. Cells that border the exterior are replaced by their convex hull. CoACD \cite{wei2022coacd} also provides a way to tune the fidelity and complexity of the decomposition.
To compare our volume decompositions with the ones of CoACD we sample the input mesh with 2M points and produce plane configurations with different complexities by varying the fitting tolerance and the minimal number of inliers per plane. 
\subsubsection{Results.} We show complexity / accuracy curves for our and CoACD's decompositions in \figref{fig:coacd}. To produce decompositions with a small number of cells ($|C_V| < 400$) our pipeline relies on a low number of input planes which are sometimes not sufficient to approximate the input well. CoACD directly relies on the input mesh and thus exhibits a higher volumetric intersection over union. However, this comes at the cost of a much higher number of convex cell facets (see \figref{subfig:coacd_b}).
For more complex and less approximate decompositions ($|C_V| > 500$) our method is both more accurate and more concise.
This is also exemplified in \figref{subfig:coacd_c}, where the top row shows two decompositions with different complexity of the \emph{Droid} model and the bottom row a decomposition of the \emph{Temple} model.
Because our method uses cutting planes that are detected on the surface of the input, the cells of our decomposition represent the geometry much better. See for example the gun of the \emph{Droid} or the stairs in the close up of the \emph{Temple}.

\subsection{Volume Decomposition with Overlapping Convexes}

\subsubsection{Experimental Setup.} A variant of the convex decomposition problem is to relax the non-overlapping constraint between convexes. To address this task, we set the merge threshold $\tau$ of \eqref{eq:simplification} to a strictly positive number, which allows us to merge neighboring cells, and extract a decomposition at any desired number of convexes.
We compare our method to BSP-Net \cite{chen2020bspnet} on the test sets of all 13 categories of ShapeNet, \ie 8762 models in total, provided by the authors of BSP-Net.
BSP-Net is a neural network based approach that learns to fit planes to input observations and assemble the planes into a set of overlapping convex polytopes. We use the auto-encoder variant of the network with weights trained (provided by the authors) on all 13 categories of ShapeNet.
The network inputs a voxel-grid of $64^3$ occupancy values per model, while we run our method on 100k points sampled on the models' surface.
To find the best trade-off between complexity and accuracy of the produced models we operate a small grid search on the ShapeNet train set to determine the merge threshold $\tau$ and the two parameters of the planar shape detection.
Alternatively, we could also merge neighboring cells until our models have the same number of convexes as the models produced by BSP-Net. However, we find that BSP-Net often does not output a  suitable number of convexes to accurately describe a models geometry.
We use the evaluation pipeline provided by the authors of BSP-Net to compute the \emph{squared} Chamfer distance (\chamfersq) and normal consistency (\normal). Note that, the pipeline computes these metrics only for surface points by using the occupancy of input voxels to determine whether a point lies inside a shape or on its surface.



\input{figures/bspnet/combined_eccv/combined}

\subsubsection{Results.} \figref{fig:bspnet} shows that our method offers both better accuracy and complexity than BSP-Net. The two visual results illustrate this quality difference with convexes capturing more details and more meaningful components of the models. Our method manages to capture thin components with single convexes whereas BSP-Net tends to regroup them into large convexes that strongly overlap. Note that the same set of parameters found by grid searching generalises well to various shapes with different feature sizes.

%% file: tables/thingi.tex
\begin{table}[t]
    \setlength{\tabcolsep}{12pt}
\centering
\caption{\textbf{Quantitative Comparison with Plane Arrangements.}
The exhaustive method is only tested on the simple group of models \emph{S} to not exceed reasonable processing time ($>300h$). 
$|C|$ refers to the average number of cells in the decompositions. $|F_S|$, CD and HD are the average number of facets of the reconstructed mesh and Chamfer and Hausdorff distances between ground truth and reconstructed mesh.
}
\vspace{\captionvertspace}
\resizebox{0.7\linewidth}{!}{%
\begin{tabular}{@{}l|lrrrrr@{}}
\toprule
\multicolumn{2}{c}{} & \multicolumn{2}{c}{\emph{Complexity}} & \multicolumn{2}{c}{\emph{Accuracy}} & \emph{Performance} \\
\midrule
\multicolumn{2}{c}{} & $|C|$              & $|F_S|$             & \chamfer &  \hausdorff   & Time     \\
\multicolumn{2}{c}{} &   &   & $(\times 10^{2})$ & $(\times 10^{2})$  & (s) \\
\midrule
\multirow{4}{*}{\rotatebox[origin=c]{90}{\emph{S}}}
&\textbf{Exhaustive}~\cite{Edelsbrunner83}  &    5690           & 890           & \bf 0.190 & \bf  1.05 & 8.03 \\
&\textbf{Adaptive}~\cite{Murali:1997:CSA}   &    324            & 70            &     0.196 &      1.32& 9.99 \\
&\textbf{KSR}~\cite{bauchet2020ksr}         &161                & 202           &     0.196 &      1.22 & 4.08 \\
&\textbf{Ours}                              &           \bf 73  & \bf 53        &     0.193     &      1.15 & \bf 3.06 \\
\midrule
\multirow{3}{*}{\rotatebox[origin=c]{90}{\emph{M}}}
&\textbf{Adaptive}~\cite{Murali:1997:CSA}  &    1760    &   267    &  \bf   0.247   &      2.36 &    97.4   \\
&\textbf{KSR}~\cite{bauchet2020ksr}        &    757     &   660    &    0.266       &      2.34 &     139   \\
&\textbf{Ours}                             &   \bf  256 & \bf  167 &    0.254       & \bf 2.32  & \bf 16.3  \\
\midrule
\multirow{2}{*}{\rotatebox[origin=c]{90}{\emph{L}}}
&\textbf{Adaptive}   \cite{Murali:1997:CSA} &         6150  &       1020    &    0.227          &     2.82  & 504   \\    
&\textbf{KSR}~\cite{bauchet2020ksr}                            &        2077   &        2313   &       0.231    &   2.79  & 1699      
\\
&\textbf{Ours}                              &      \bf  705 &  \bf  478     & \bf  0.224        & \bf 2.78  & \bf 68.1  
\\
\bottomrule
\end{tabular}
}
\label{tab:partition}
\end{table}

%% file: figures/thingi/thingi_figure.tex
\begin{figure}[t]
\newcommand{\myfontsize}{\footnotesize}
	\definetrim{mytrim1}{150 110 150 150}
	\definetrim{mytrim2}{80 20 120 90}
	\definetrim{mytrim3}{100 50 100 50}
	\newcommand{\mywidth}{0.20\columnwidth}
	\newcommand{\mywidtht}{0.20\columnwidth}
\centering
\begin{tabular}{@{}c@{}c@{}c@{}c@{}c@{}}
\includegraphics[width=\mywidtht,mytrim1]{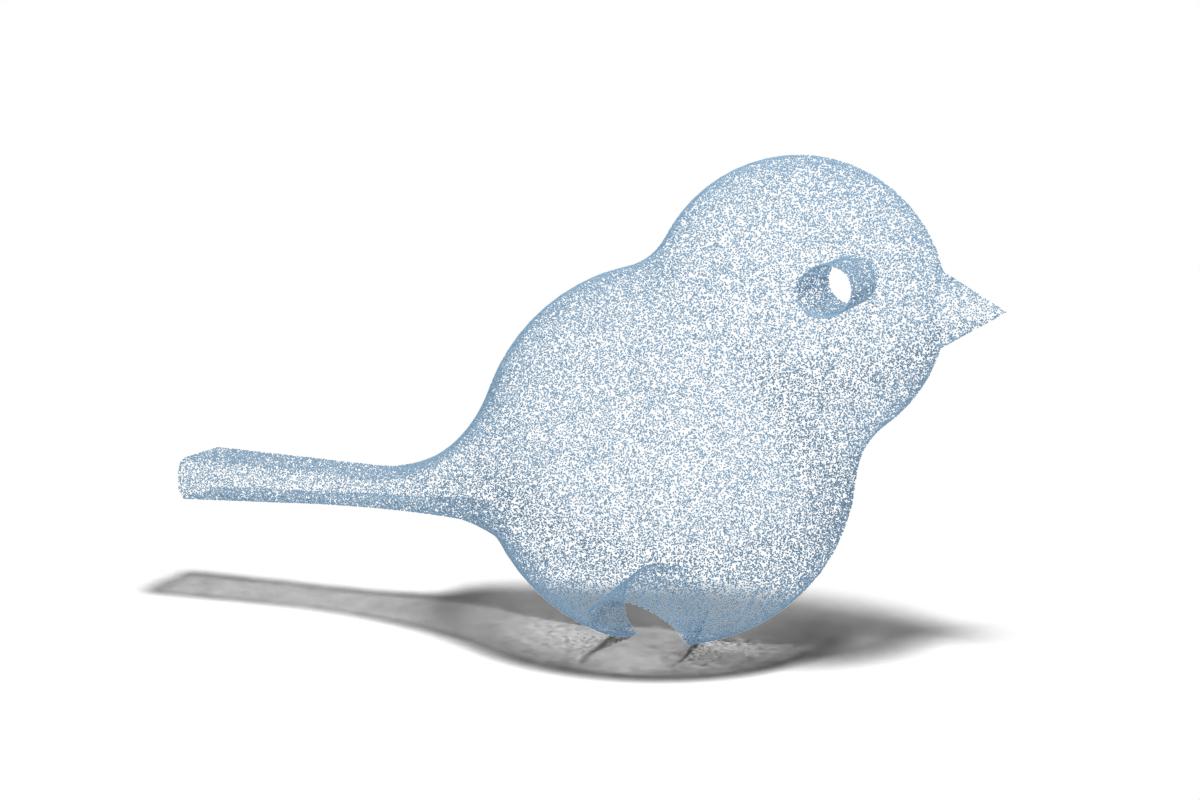}
&
\includegraphics[width=\mywidtht,mytrim2]{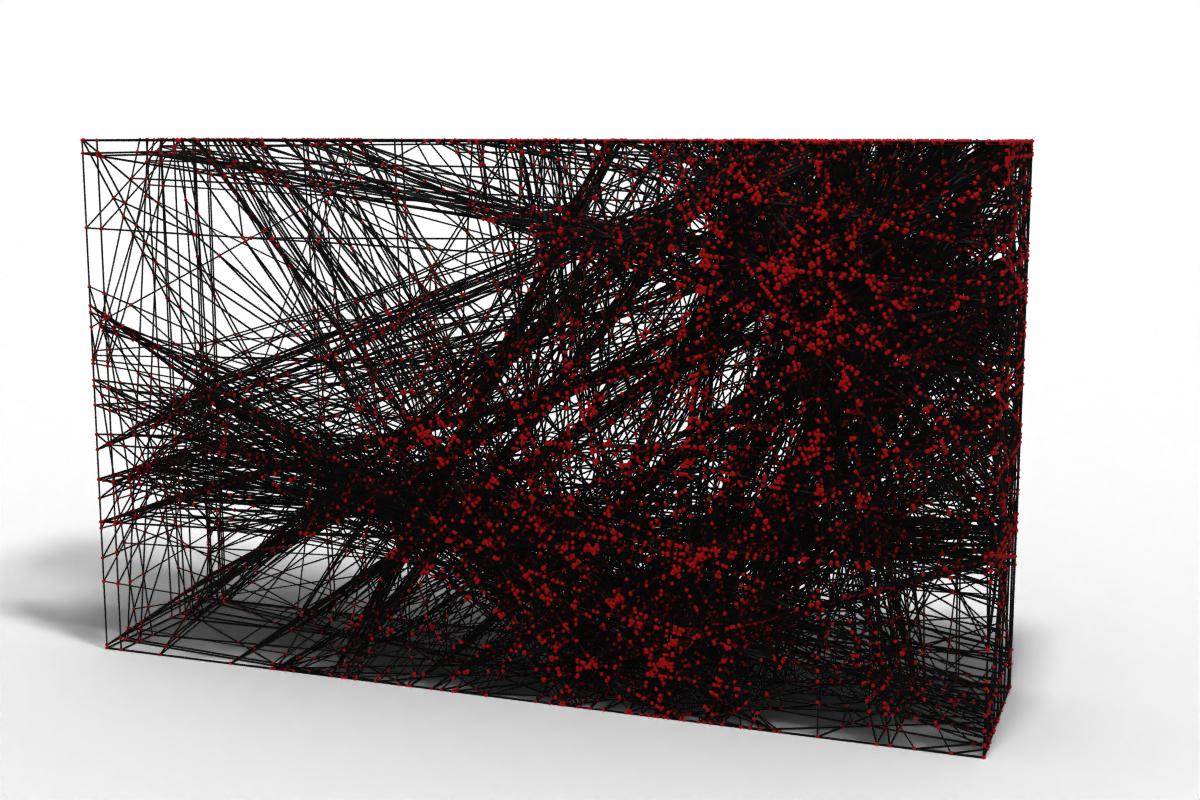}
&
\includegraphics[width=\mywidtht,mytrim2]{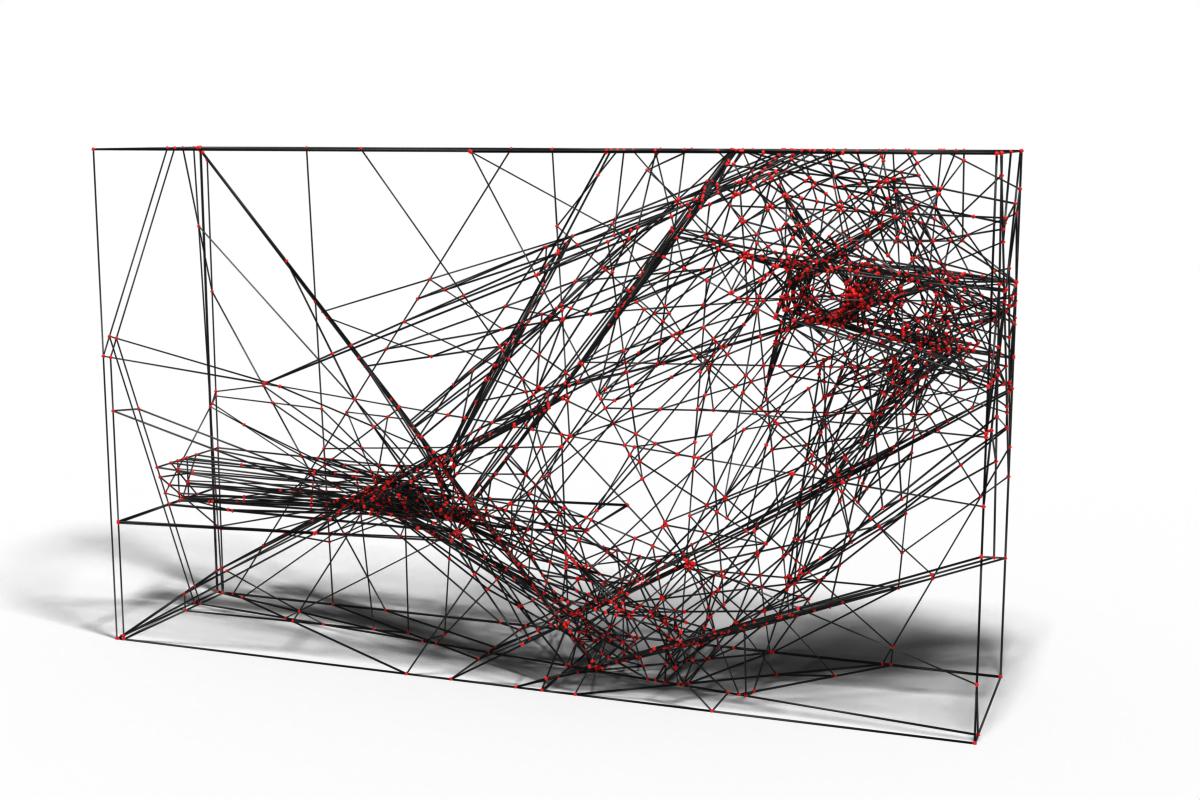}
&
\includegraphics[width=\mywidtht,mytrim2]{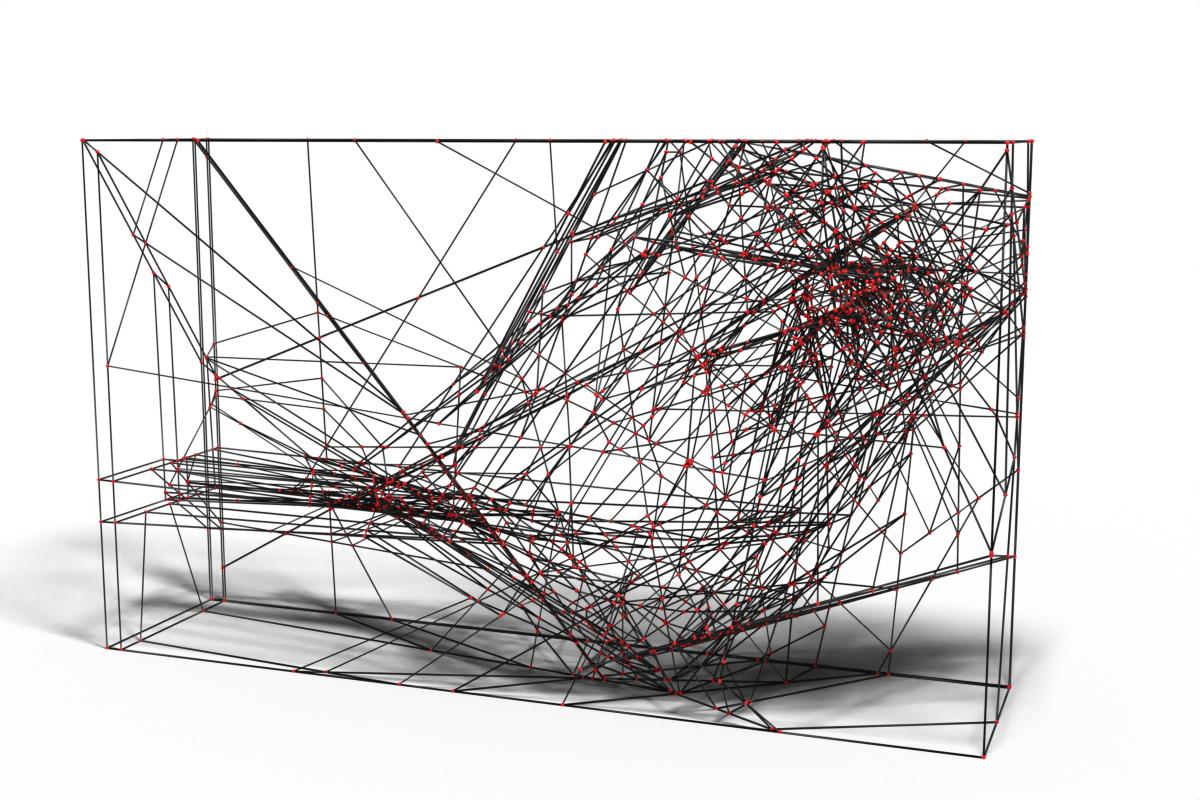}
&
\includegraphics[width=\mywidtht,mytrim2]{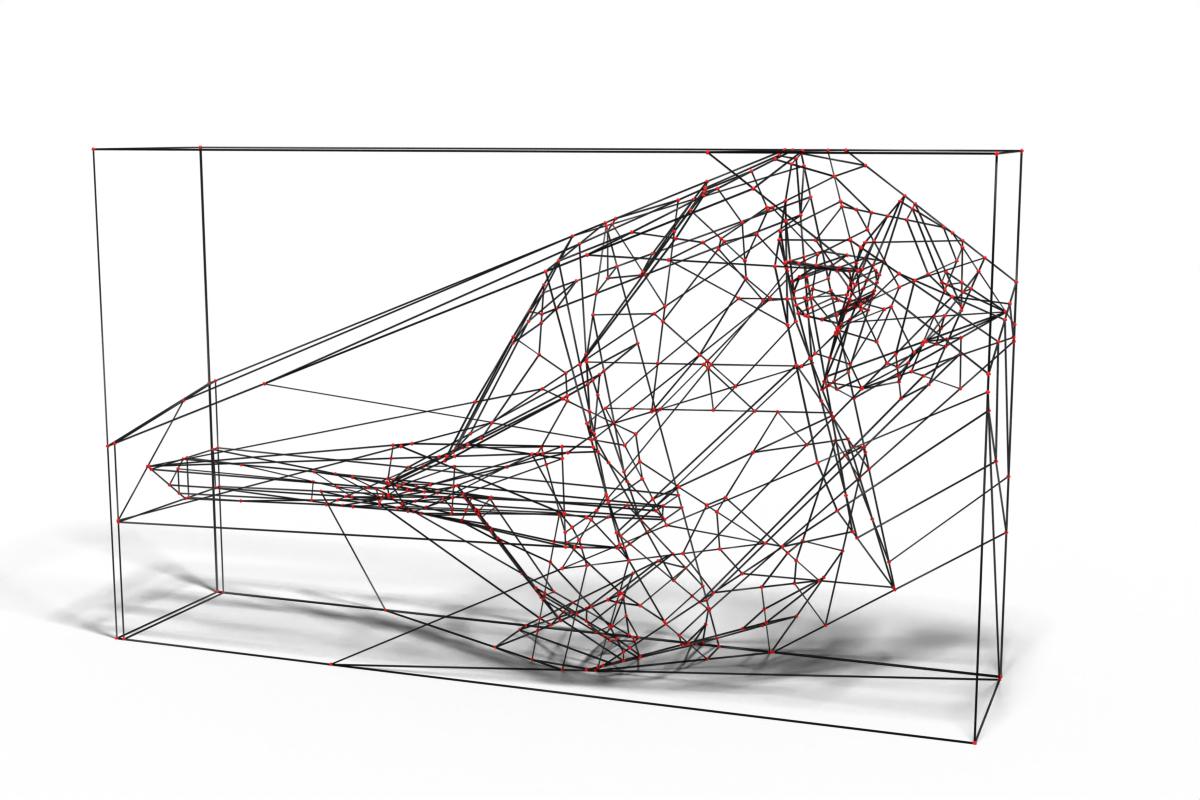}
\\
\multicolumn{5}{l}{\vspace{-0.5cm}\myfontsize Point cloud}
\\
\includegraphics[width=\mywidtht,mytrim1]{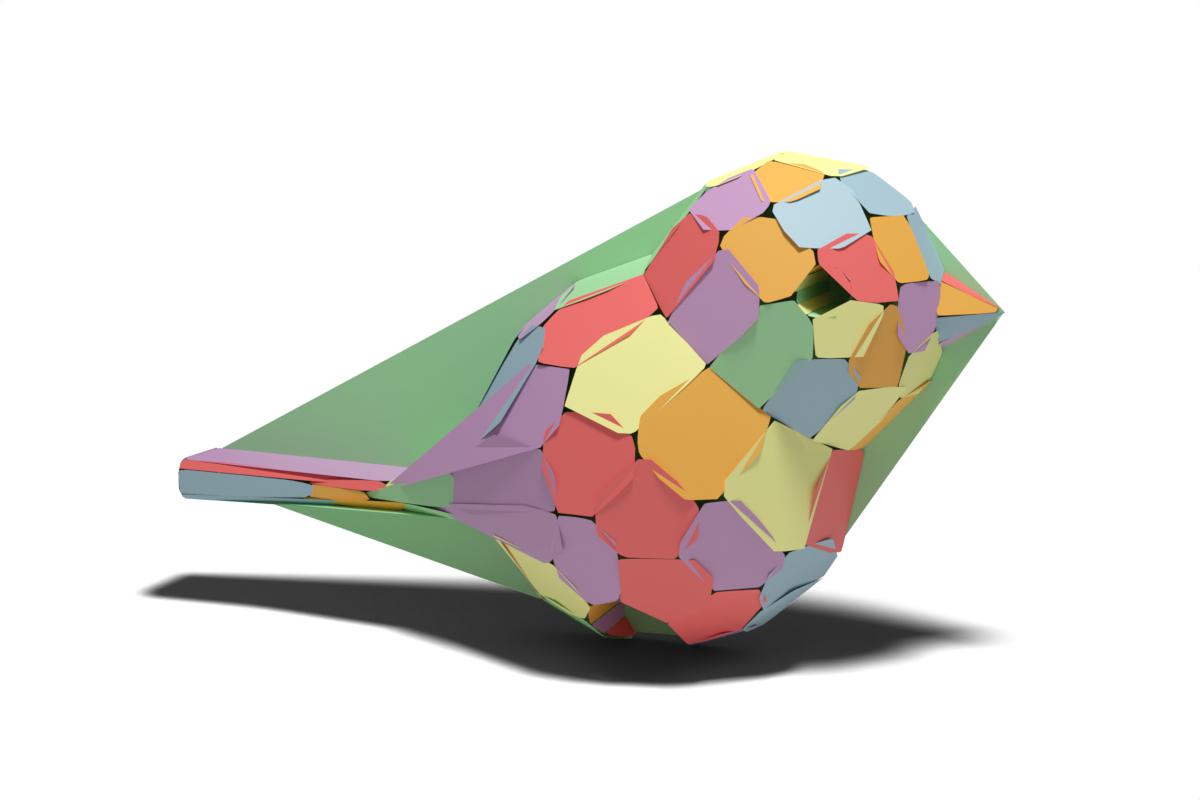}
&
\includegraphics[width=\mywidtht,mytrim1]{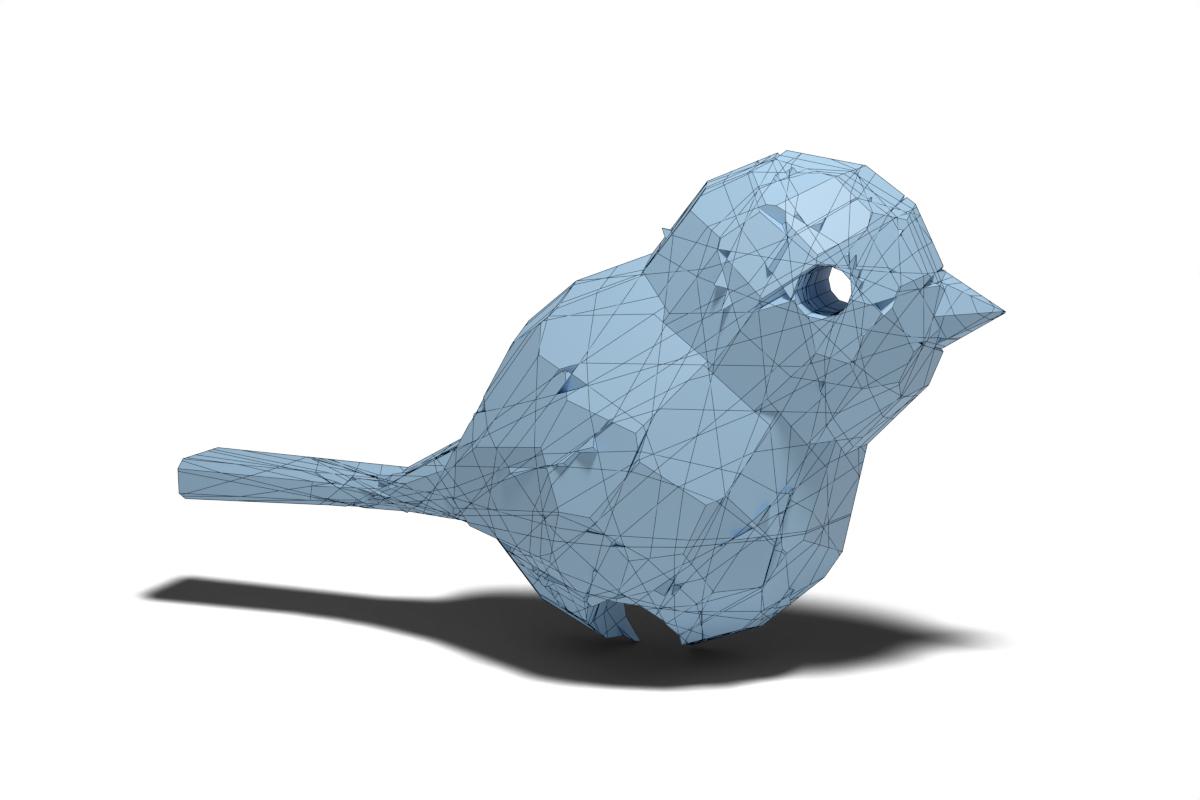}
&
\includegraphics[width=\mywidtht,mytrim1]{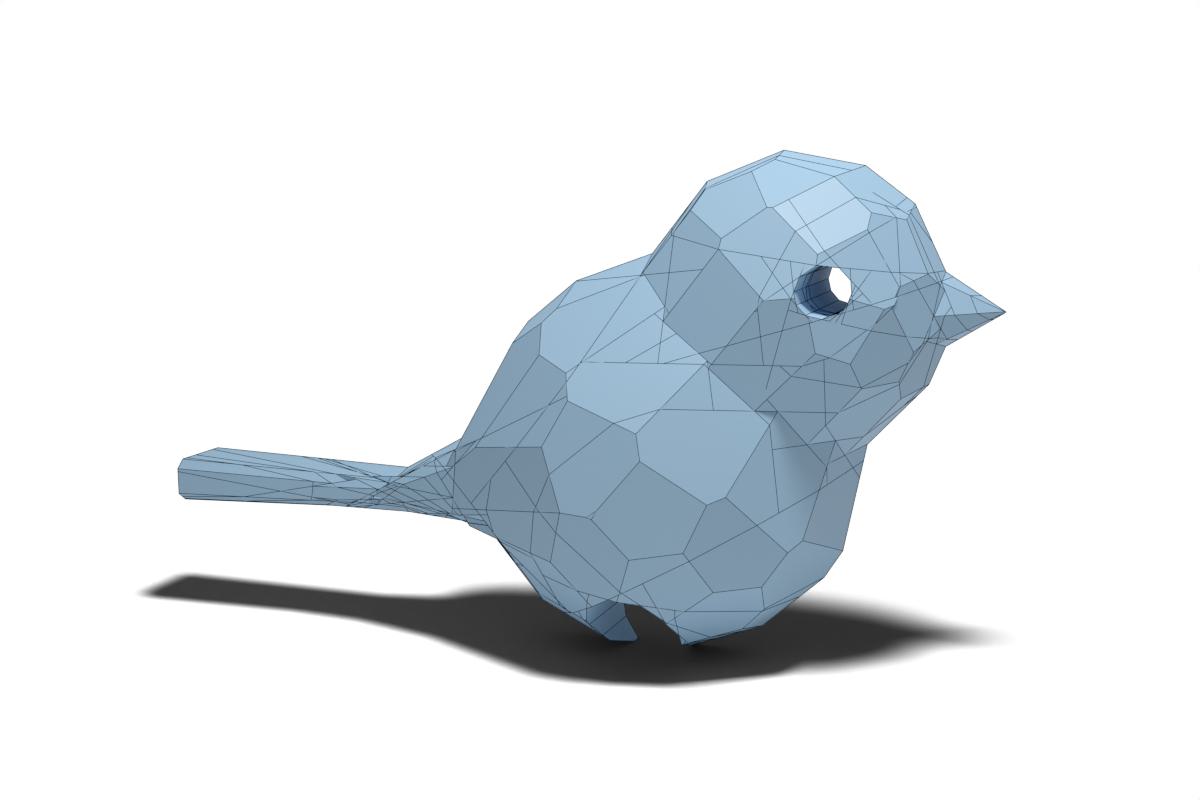}
&
\includegraphics[width=\mywidtht,mytrim1]{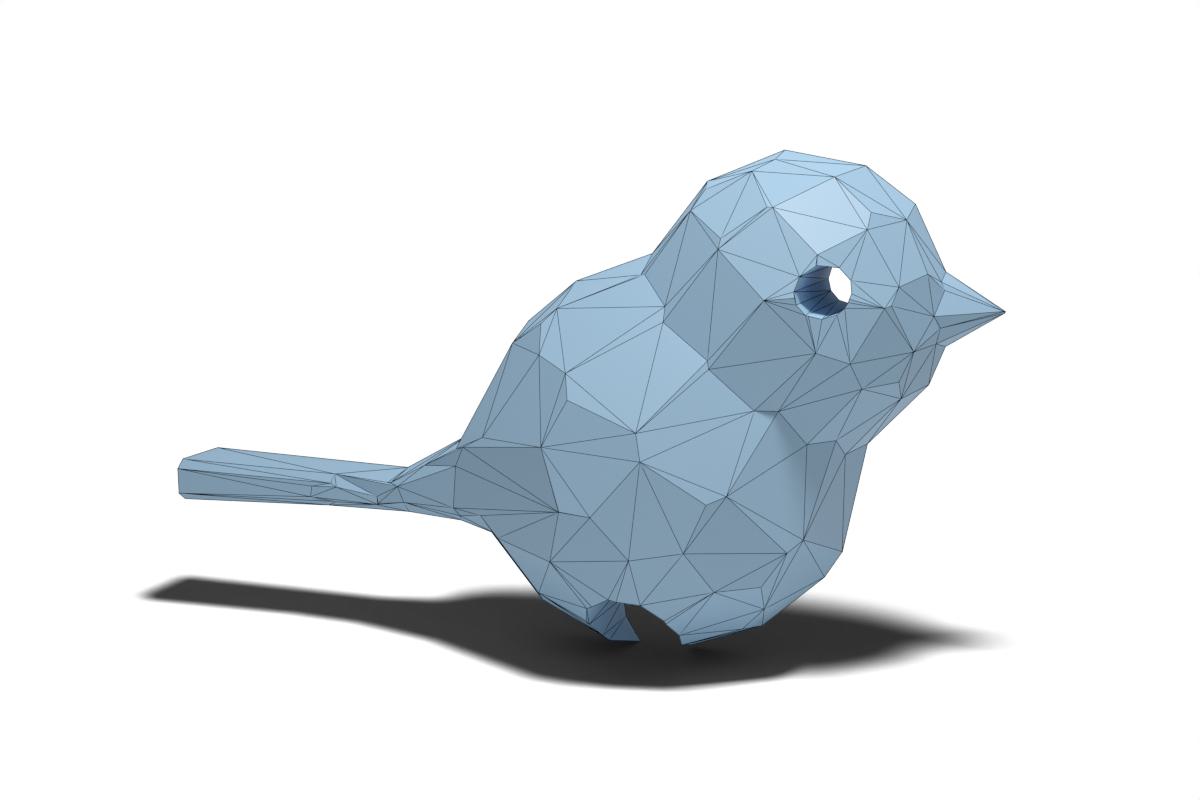}
&
\includegraphics[width=\mywidtht,mytrim1]{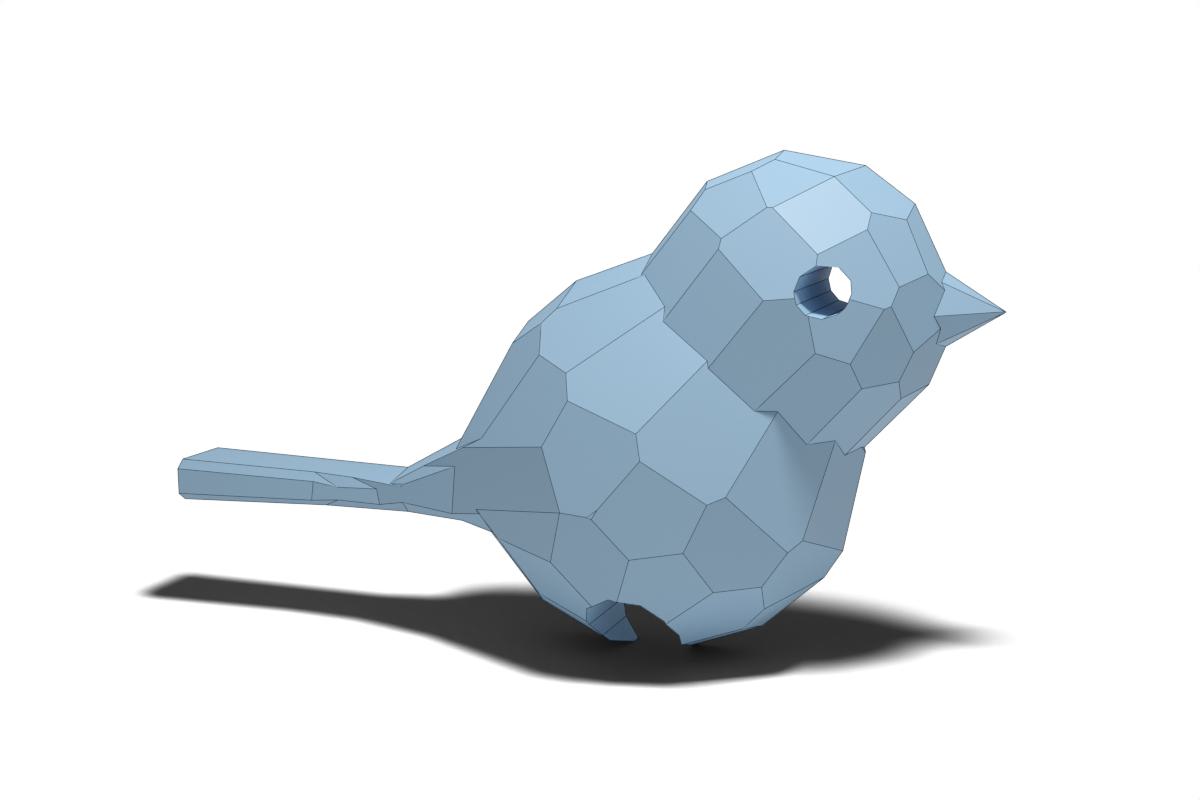}
\\
\myfontsize PC \& planes & \myfontsize Exhaustive~\cite{Edelsbrunner83} & \myfontsize Adaptive~\cite{Murali:1997:CSA} &  \myfontsize KSR~\cite{bauchet2020ksr} &   \myfontsize Ours
\end{tabular}
\vspace{\captionvertspace}
\caption{\textbf{Comparison with Plane Arrangements.}
Our algoritm builds a more concise plane arrangement than existing mechanisms (top row), leading to a polyon mesh with fewer facets (bottom row).
}
\label{fig:thingi}
\end{figure}

%% file: figures/runtime/runtime_figure.tex
\begin{figure}[t]

    \centering
\includegraphics[width=0.7\columnwidth,trim=18 0 20 20,clip]{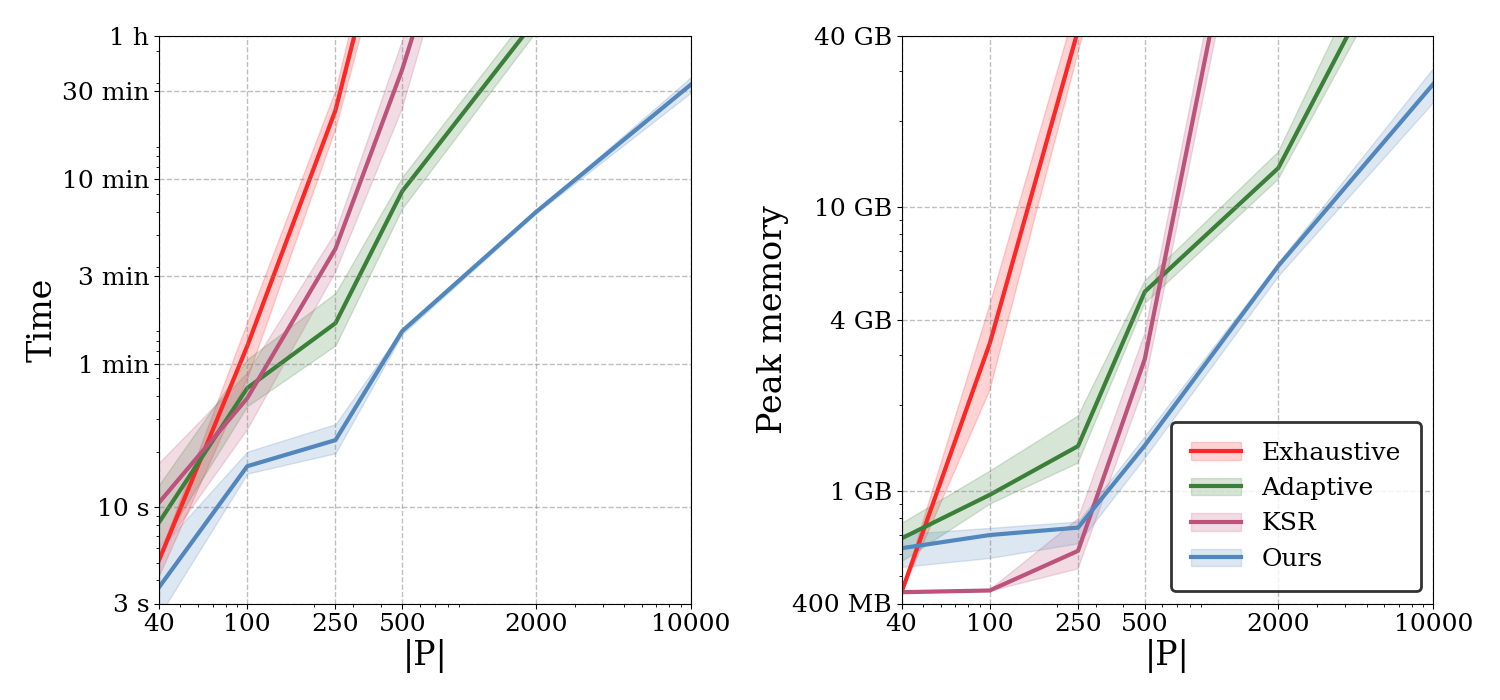}

 \vspace{\captionvertspace}
\caption{\textbf{Scalability.} Average runtime (left) and memory peak (right) in function of the number of input planar shapes for different construction mechanisms. The transparent band around each curve indicates the minimal and maximal values measured on various models. Our algorithm offers the best performance and can process 10k planar shapes in around 30 minutes without exceeding the memory consumption of a standard computer. It also exhibits a better stability than other methods whose variation bands are thicker.}
\label{fig:runtime}

\end{figure}

%% file: tables/ablation.tex
\begin{table}[t]
    \setlength{\tabcolsep}{12pt}
\centering
\caption{\textbf{Ablation Study.}
Alternative schemes to the use of inlier points and for the sorting of splitting operations, as well as the deactivation of the remeshing step are evaluated from the complex group of the Thingi10k models. 
}
\vspace{\captionvertspace}
\resizebox{0.8\columnwidth}{!}{%
\begin{tabular}{@{}lrrrrr@{}}
\toprule
 & \multicolumn{2}{c}{\emph{Complexity}} & \multicolumn{2}{c}{\emph{Accuracy}} & \emph{Performance} \\
\midrule
& $|C|$         & $|F_S|$         & \chamfer          &  \hausdorff        &  Time \\
&               &               & $(\times 10^{2})$ & $(\times 10^{2})$ &  (s)  \\
\midrule
\textbf{Ours}                & 705           & \bf 478        &  \bf 0.224         &  2.78              & 68.1  \\
\midrule
\multicolumn{6}{l}{\emph{Insertion order}} \\
\midrule
\textbf{Convex hull area (high to low)}  & 1010      & 549           & 0.231              & 3.10             &   125  \\
\textbf{\boldmath$\argmax_p(|L_p| |R_p|)$ only}   &  920  &  491  &    0.225           &     2.87         &  \bf  57.4  \\
\midrule
\multicolumn{6}{l}{\emph{Use of inlier points}} \\
\midrule
\textbf{Convex hull vertices only}     & \bf 653      & 539           & 0.302             &  5.25             &  58.9  \\
\textbf{Points sampled on convex hull}      & 839           &  \bf 478            & 0.226             & \bf 2.73             & 66.1  \\
\midrule
\multicolumn{6}{l}{\emph{Remeshing and simplification}} \\
\midrule
\textbf{Without facet aggregation}               &      705  &   557         &  \bf 0.224             &  2.78              &   68.5  \\
\textbf{Without cell aggregation}          & 846      &  \bf 478       & \bf 0.224              &  2.78             &   73.5  \\
\bottomrule
\end{tabular}
}
\label{tab:ablation}
\end{table}

%% file: figures/robust/combined_eccv/combined.tex
\begin{figure}
    
\begin{tabular}{@{}cc@{}}

\begin{subfigure}{0.28\linewidth}
    \input{figures/robust/combined_eccv/table.tex}
    \caption{Quantitative}
    \label{subfig:robust_a}
\end{subfigure}
&
\begin{subfigure}{0.7\linewidth}
    \input{figures/robust/combined_eccv/figure.tex}
    \caption{Qualitative}
    \label{subfig:robust_b}
\end{subfigure}
\end{tabular}
\vspace{-0.2cm}
\caption{\textbf{Surface Mesh Simplification.} We compare RLPM and our pipeline on models from Thingi10k. 
We triangulate our output and perform edge collapse based on QEM s.t. OursTri and RLPM have the same number of triangles. \Subref{subfig:robust_a} The runtimes of the methods, number of vertices $|V_S|$ and facets $|F_S|$ of the surface meshes (\ie triangles for RLPM and OursTri, and polygons for Ours), and Chamfer (CD) and Hausdorff distance (HD) and normal consistency (NC) between ground truth and reconstruction. \Subref{subfig:robust_b} The reconstructions of the \emph{Tower of Pi} from the dataset. Note how both, our polygon and our triangle mesh is much more detailed compare to the one of RLPM. 
}
\label{fig:robust}
\end{figure}

%% file: figures/robust/combined_eccv/table.tex
\resizebox{\linewidth}{!}{%
\begin{tabular}{@{}c@{}}
    \setlength{\tabcolsep}{7pt}
\begin{tabular}{@{}l@{}rrr}
    \toprule
    & \emph{Perform.} & \multicolumn{2}{c}{\emph{Complexity}}  \\
    \midrule
    & Time & $|V_S|$ & $|F_S|$  \\
    & (s) &  &  \\
    \midrule
    \textbf{RLPM~\cite{chen2023robust}} & 503 & 899 & 1969   \\
    \textbf{Ours} & \bf 410  & 2051 &  2045  \\
    \textbf{OursTri}  & 411 & \bf 879 & \bf 1936 \\
    \bottomrule
\end{tabular}
\\[1.5cm]
\setlength{\tabcolsep}{9pt}
\begin{tabular}{rrr@{}}
    \toprule
    \multicolumn{3}{c}{\emph{Accuracy}}  \\
    \midrule
    \chamfer~$\downarrow$ & \hausdorff~$\downarrow$ & \normal~$\uparrow$ \\
    ($\times 10^{2}$) & ($\times 10^{2}$)   \\
    \midrule
    0.269 & 2.48 & 0.923   \\
    \bf 0.218 &  \bf 2.10 & \bf 0.941  \\
     0.244 & 2.32 & 0.924  \\
    \bottomrule
\end{tabular}
\end{tabular}}

%% file: figures/robust/combined_eccv/figure.tex
\definetrim{mytrim1}{50 10 50 0}
\newcommand{\mywidth}{0.25\columnwidth}
\newcommand{\mywidthd}{\columnwidth}
\newcommand{\myfontsize}{\footnotesize}

\tikzstyle{closeup} = [
    opacity=1.0,          
    height=0.9cm,         
    width=0.9cm,          
    magnification=2.5,
    connect spies, yellow  
  ]
  \tikzstyle{largewindow} = [circle, yellow, line width=0.35mm]
  \tikzstyle{smallwindow} = [circle, yellow, line width=0.25mm]

\begin{tabular}{@{}c@{}c@{}c@{}c@{}}

\begin{subfigure}{\mywidth}
    \begin{tikzpicture}[spy using outlines={every spy on node/.append style={smallwindow}}]

      \coordinate (ai) at (-0.78,1.02);
      \coordinate (aj) at (0.5,1.6);

      \node[anchor=south] (FigA) at (0,0){%
      \includegraphics[width=\mywidthd,mytrim1]{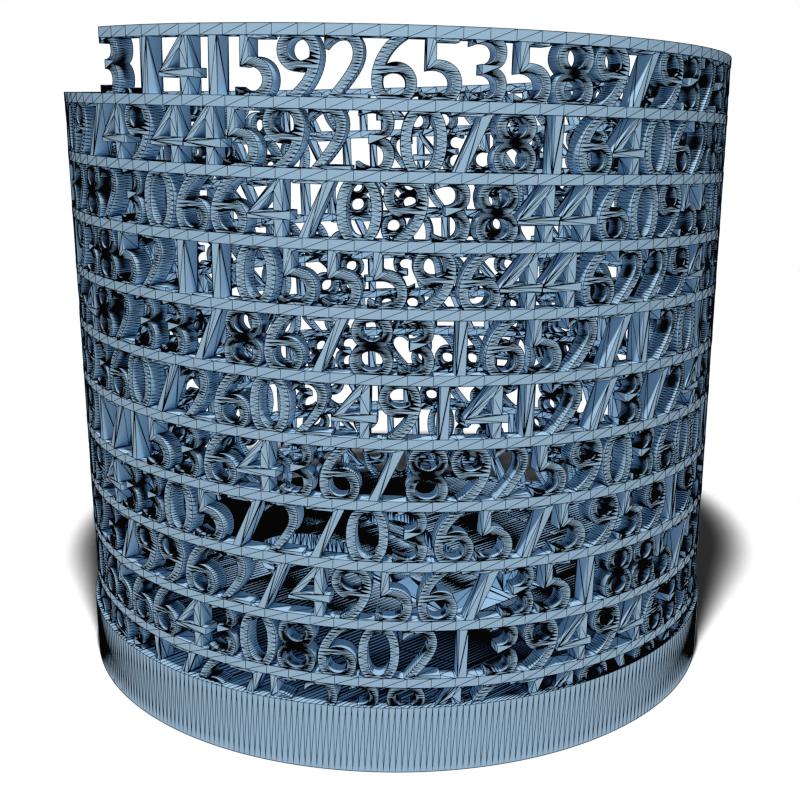}};
      \spy [closeup] on ($(FigA)+(ai)$) 
          in node[largewindow] at ($(FigA)+(aj)$);
    \end{tikzpicture}
\end{subfigure}
&
\begin{subfigure}{\mywidth}
    \begin{tikzpicture}[spy using outlines={every spy on node/.append style={smallwindow}}]

      \node[anchor=south] (FigA) at (0,0){%
      \includegraphics[width=\mywidthd,mytrim1]{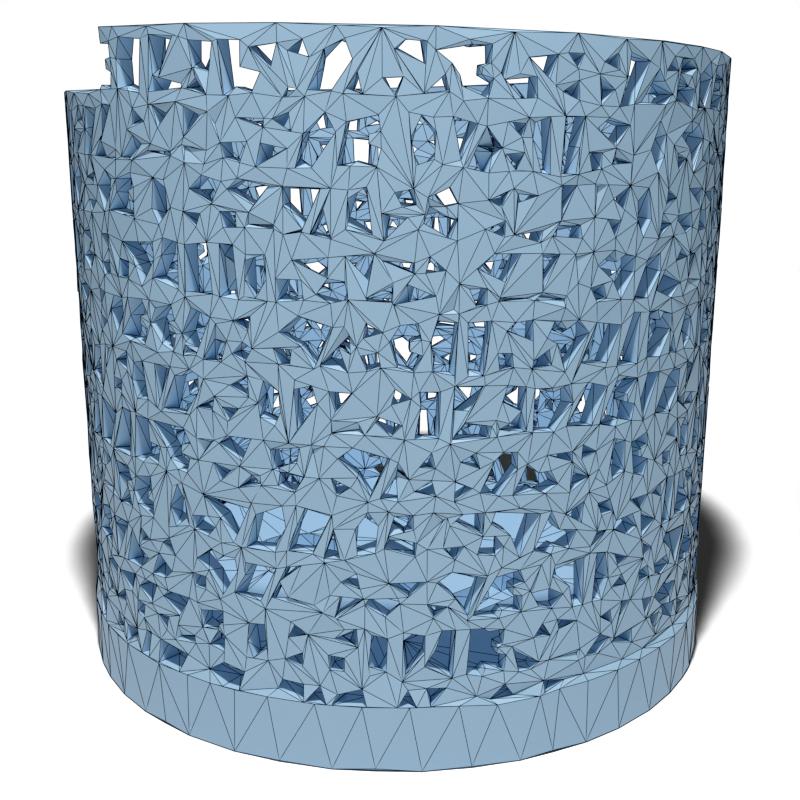}};
      \spy [closeup] on ($(FigA)+(ai)$) 
          in node[largewindow] at ($(FigA)+(aj)$);
    \end{tikzpicture}
\end{subfigure}
&
\begin{subfigure}{\mywidth}
    \begin{tikzpicture}[spy using outlines={every spy on node/.append style={smallwindow}}]

      \node[anchor=south] (FigA) at (0,0){%
      \includegraphics[width=\mywidthd,mytrim1]{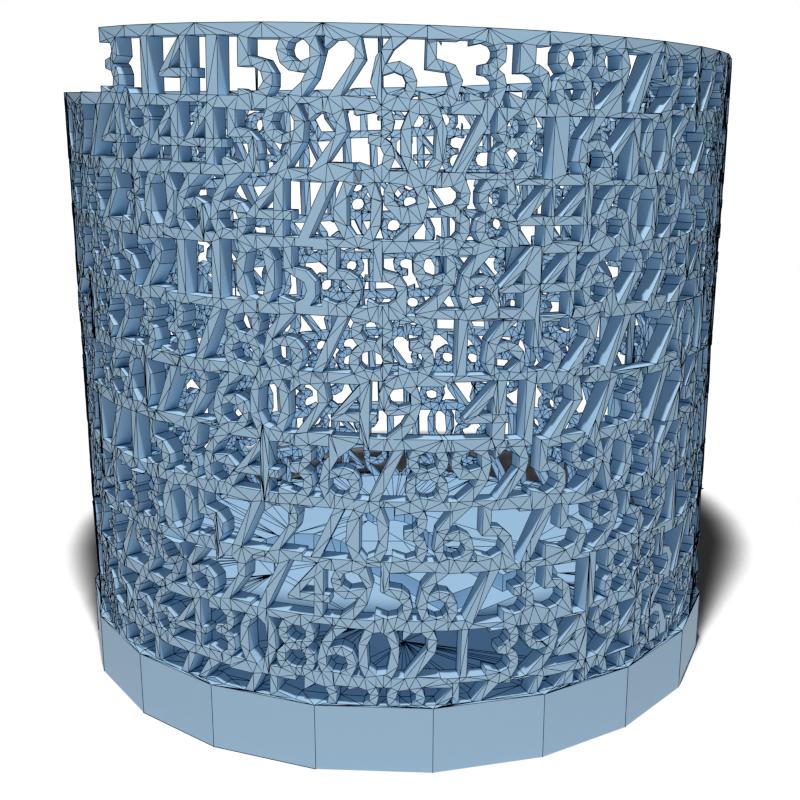}};
      \spy [closeup] on ($(FigA)+(ai)$) 
          in node[largewindow] at ($(FigA)+(aj)$);
    \end{tikzpicture}
\end{subfigure}
&
\begin{subfigure}{\mywidth}
    \begin{tikzpicture}[spy using outlines={every spy on node/.append style={smallwindow}}]

      \node[anchor=south] (FigA) at (0,0){%
      \includegraphics[width=\mywidthd,mytrim1]{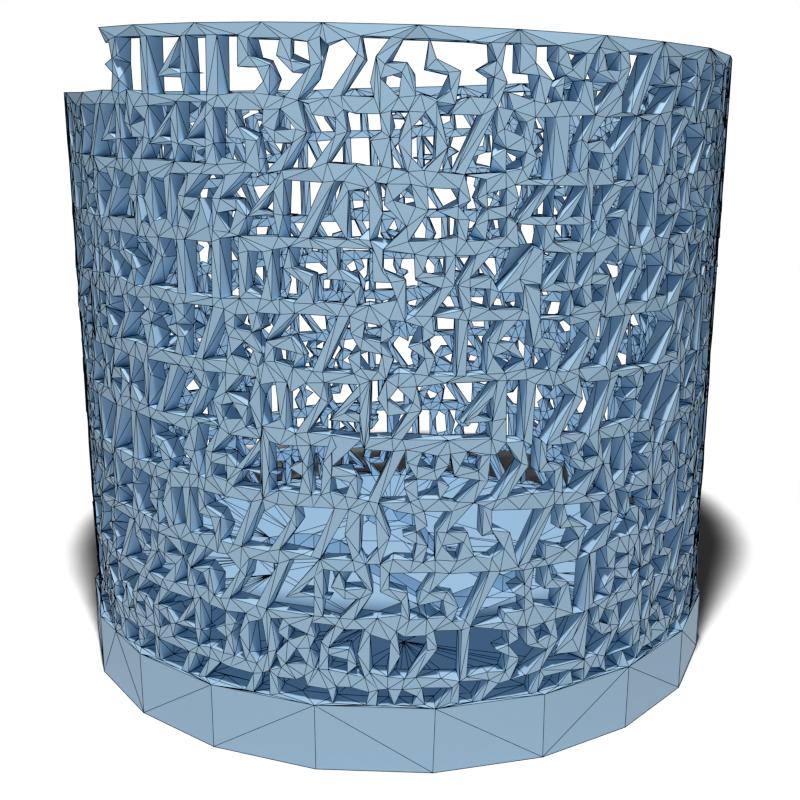}};
      \spy [closeup] on ($(FigA)+(ai)$) 
          in node[largewindow] at ($(FigA)+(aj)$);
    \end{tikzpicture}
\end{subfigure}
\vspace{\captionvertspace}
\\
\myfontsize Input& \myfontsize RLPM~\cite{chen2023robust} 
& \myfontsize Ours & \myfontsize OursTri
\end{tabular}

%% file: figures/coacd/coacd_combined_eccv.tex
\begin{figure}[ht]
    \begin{tabular}{@{}cc@{}}
        \centering
        \begin{subfigure}{0.25\linewidth}
            \includegraphics[width=\linewidth,trim=18 0 2 10,clip]{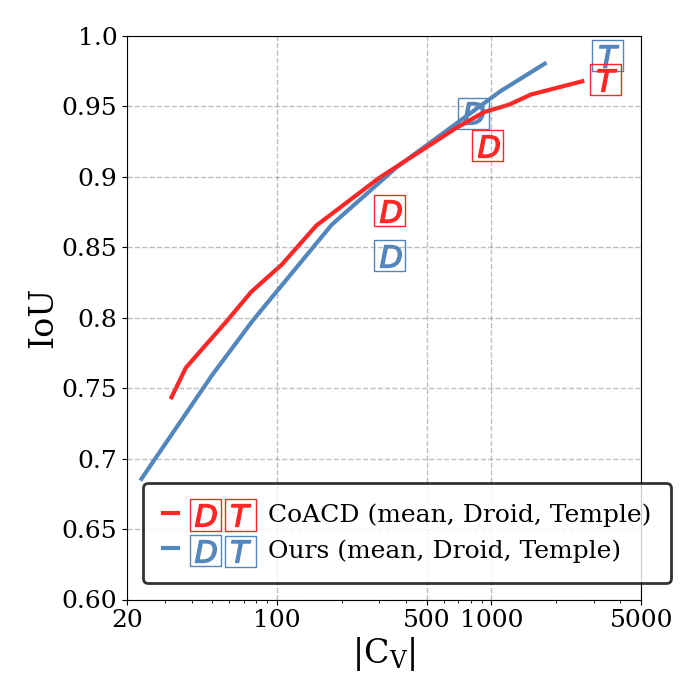}
        \caption{Cells}
        \label{subfig:coacd_a}
        \end{subfigure}
        &
        \multirow[vpos]{2}{*}{
        \centering
        \begin{subfigure}{0.74\linewidth}
            \addtocounter{subfigure}{1}
            \vspace{-8\baselineskip}
            \input{figures/coacd/coacd_figure_eccv}
        \caption{Qualitative}
        \label{subfig:coacd_c}
        \end{subfigure}}
        \\
        \begin{subfigure}{0.25\linewidth}
            \addtocounter{subfigure}{-2}
            \includegraphics[width=\linewidth,trim=18 0 2 10,clip]{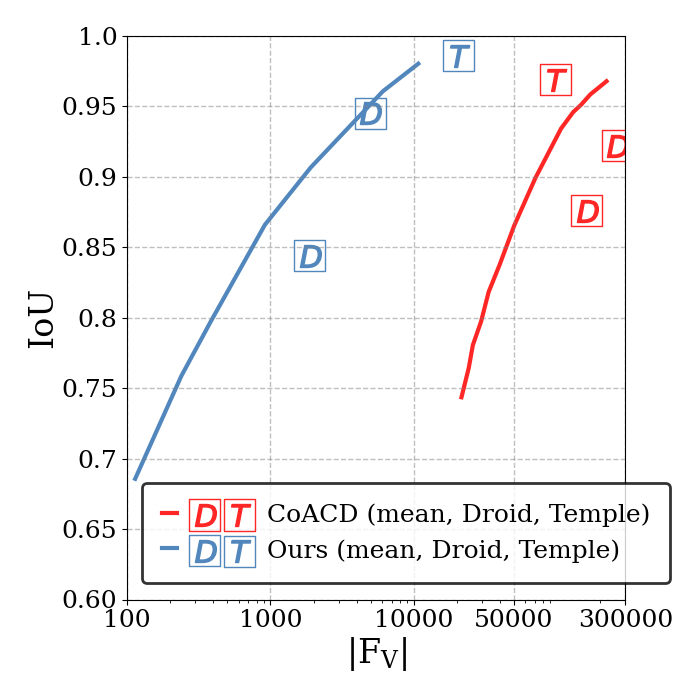}
        \caption{Facets}
        \label{subfig:coacd_b}
        \end{subfigure}
    \end{tabular}
    \vspace{\captionvertspace}
    \caption{\textbf{Volume Decomposition with Non-Overlapping Convexes.} 
    We plot the volumetric IoU in function of the number of cells $|C_V|$ \Subref{subfig:coacd_a} and facets $|F_V|$ \Subref{subfig:coacd_b}.
    Our algorithm offers a trade-off between IoU and number of convexes similar to the one of CoACD \cite{wei2022coacd}, but with convexes with fewer facets.
    In \Subref{subfig:coacd_c}, the top row shows two decompositions for both methods of the \emph{Droid} with different levels of complexity. The bottom row shows a decomposition of the \emph{Temple}.
For a similar number of convexes (colored cells), our algorithm captures the geometry better (see stairs on the bottom close-up), with fewer facets (see edges on the top close up).}
\label{fig:coacd}
\end{figure}

%% file: figures/coacd/coacd_figure_eccv.tex
\definetrim{temple}{80 0 130 0}
\definetrim{templeclose}{230 20 200 310}
\newcommand{\temple}{0.33\columnwidth}
\definetrim{droid}{0 0 10 0}
\definetrim{droidclose}{0 250 450 550}
\newcommand{\droid}{0.33\columnwidth}
\newcommand{\myfontsize}{\footnotesize}
\begin{tabular}{@{}c@{}c@{}c@{}}
\begin{overpic}[width=\droid,droid]
{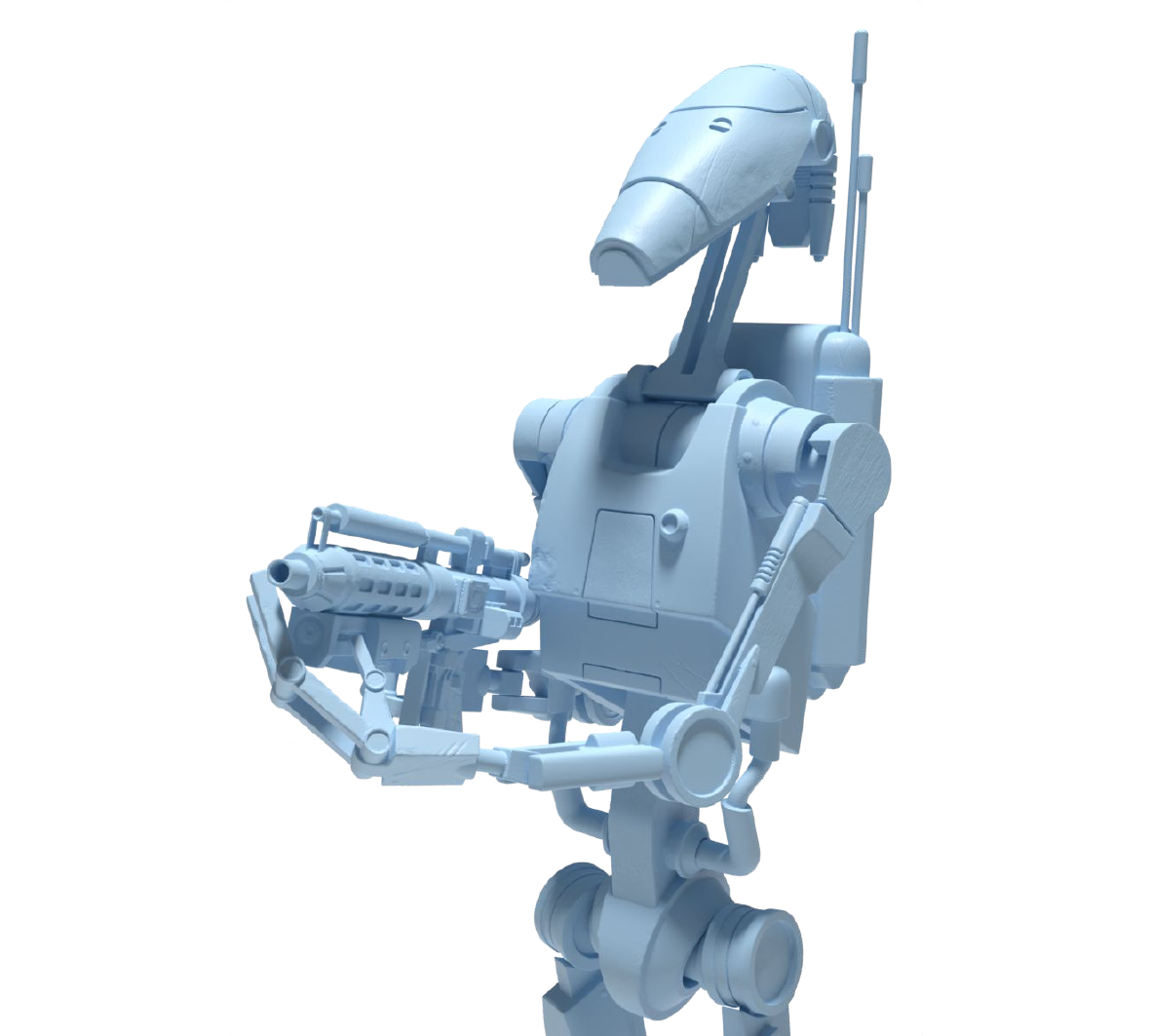}
\end{overpic}
&
\begin{overpic}[width=\droid,droid]
{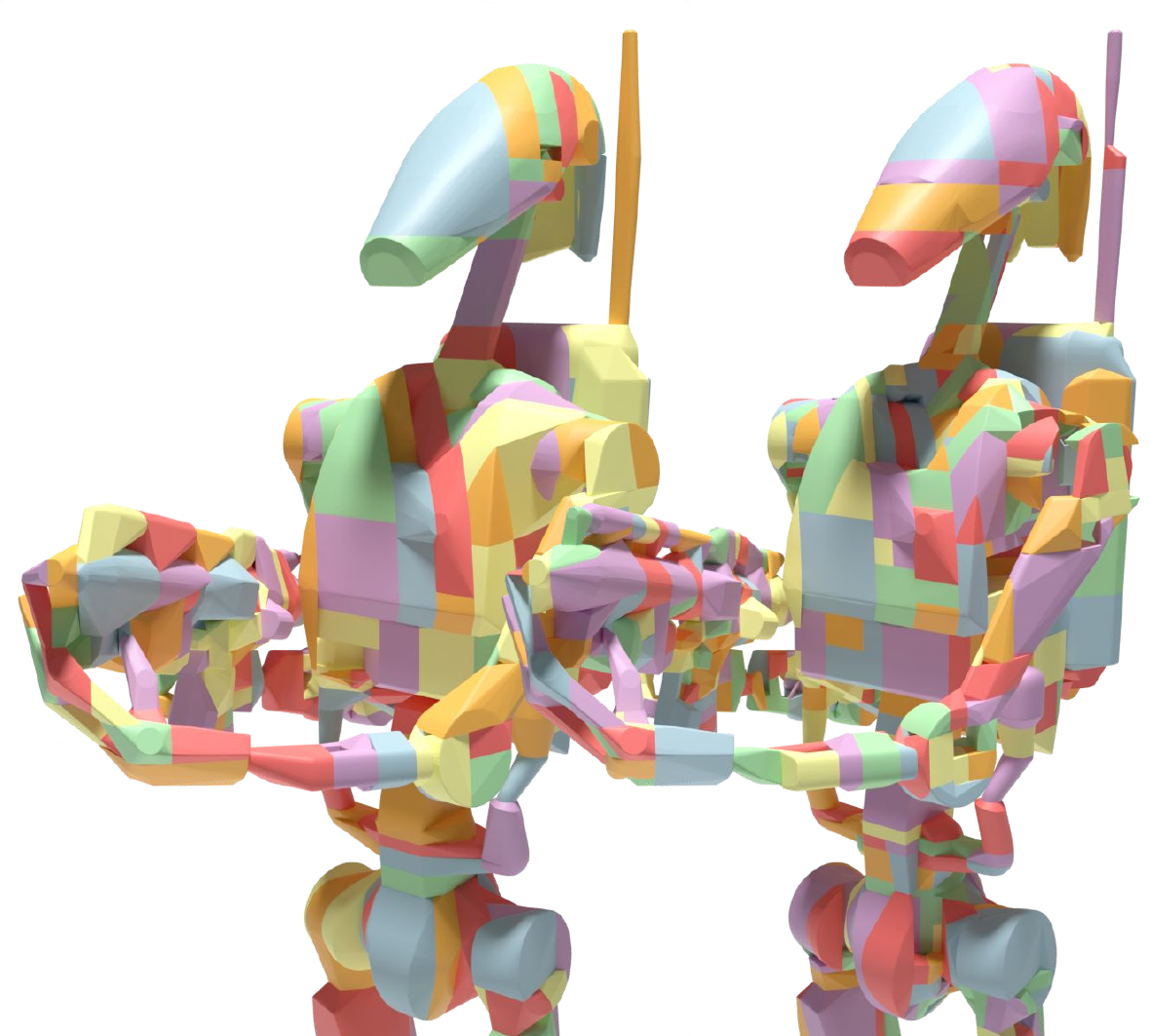}
\put(30,600){\color{yellow}\linethickness{0.35mm}%
\frame{\includegraphics[scale=.06,droidclose]{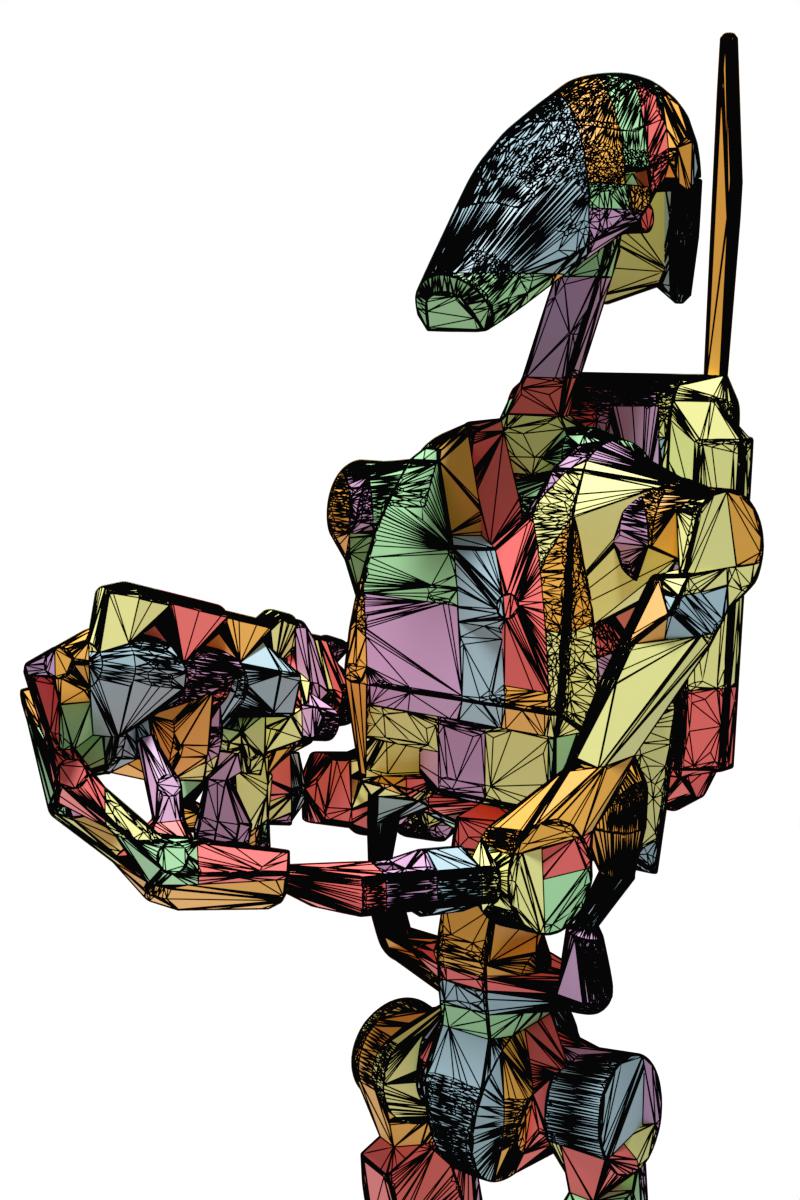}}}
\end{overpic}
&
\begin{overpic}[width=\droid,droid]
{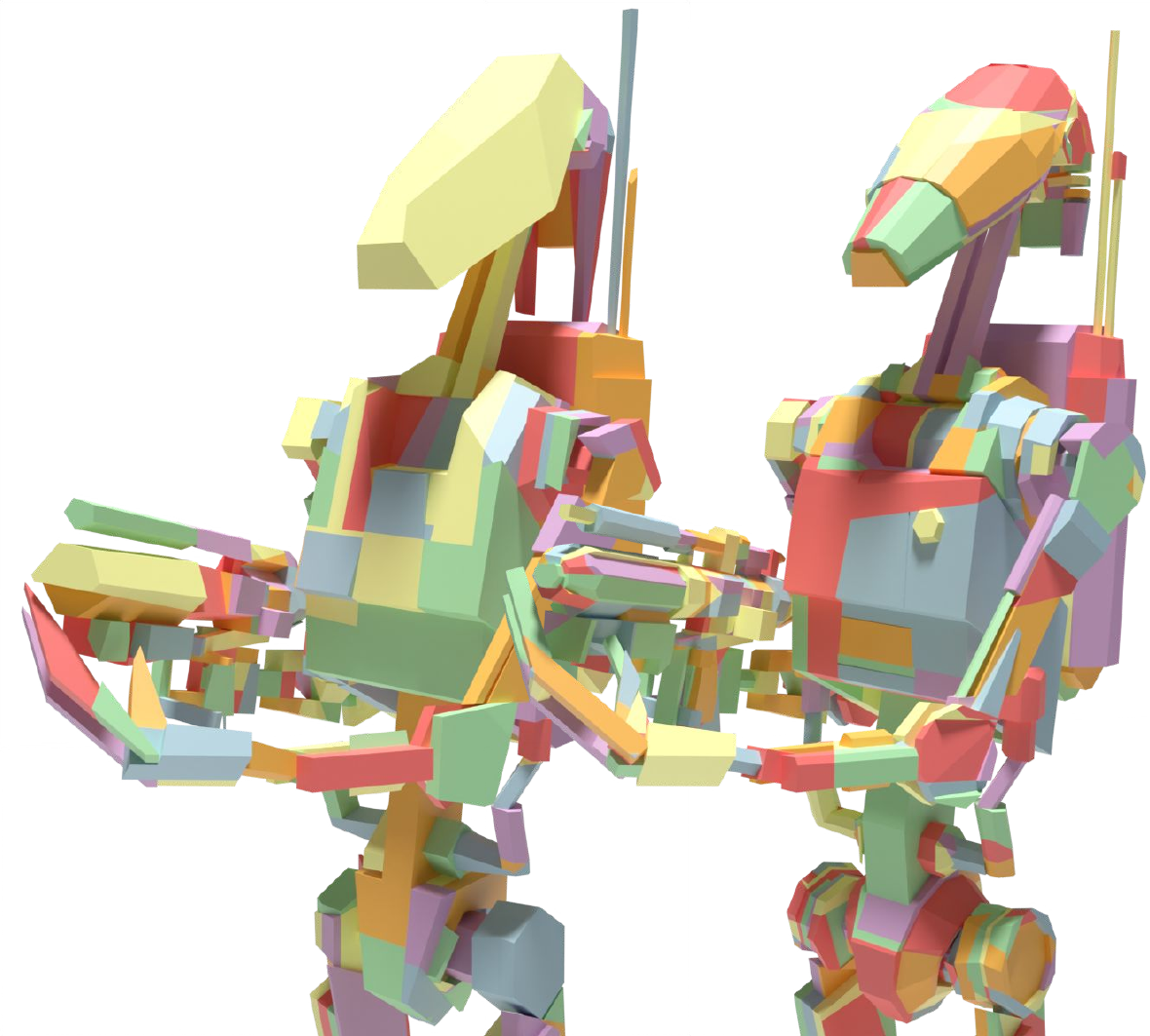}
\put(30,600){\color{yellow}\linethickness{0.35mm}%
\frame{\includegraphics[scale=.06,droidclose]{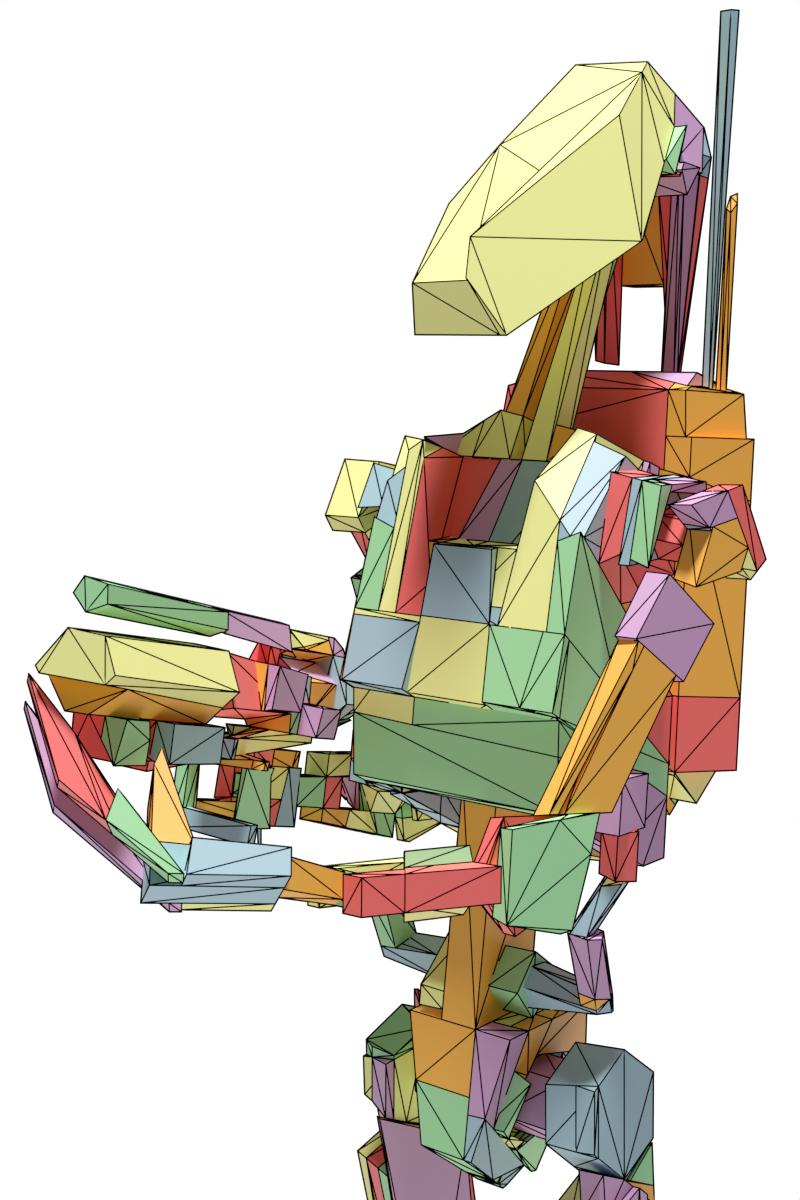}}}
\end{overpic}
\\
\begin{overpic}[width=\temple,temple]
{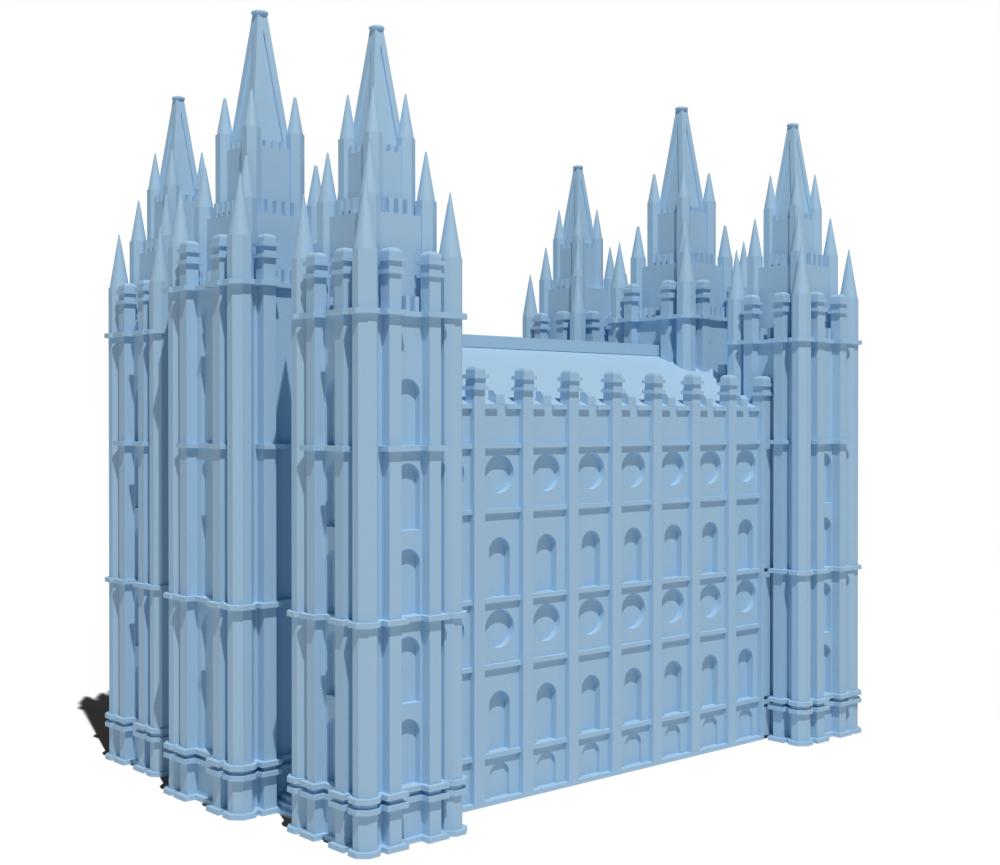}
\put(600,0){\color{yellow}\linethickness{0.35mm}%
\frame{\includegraphics[scale=.045,templeclose]{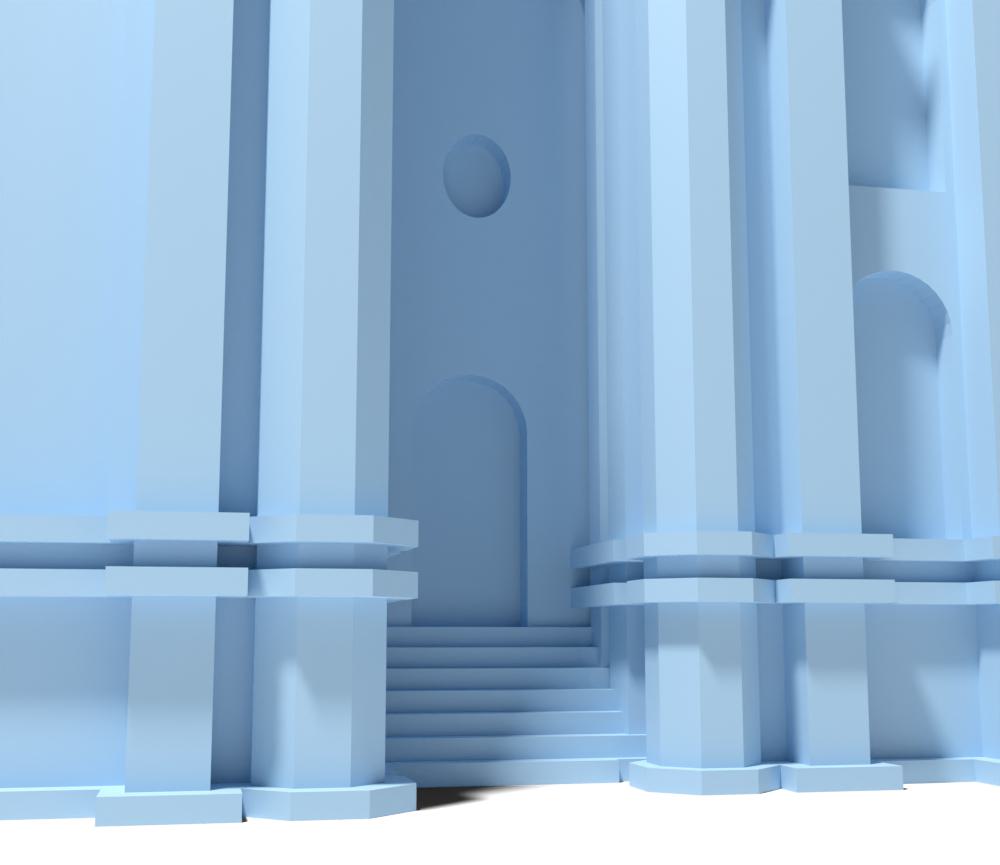}}}
\end{overpic}
&
\begin{overpic}[width=\temple,temple]
{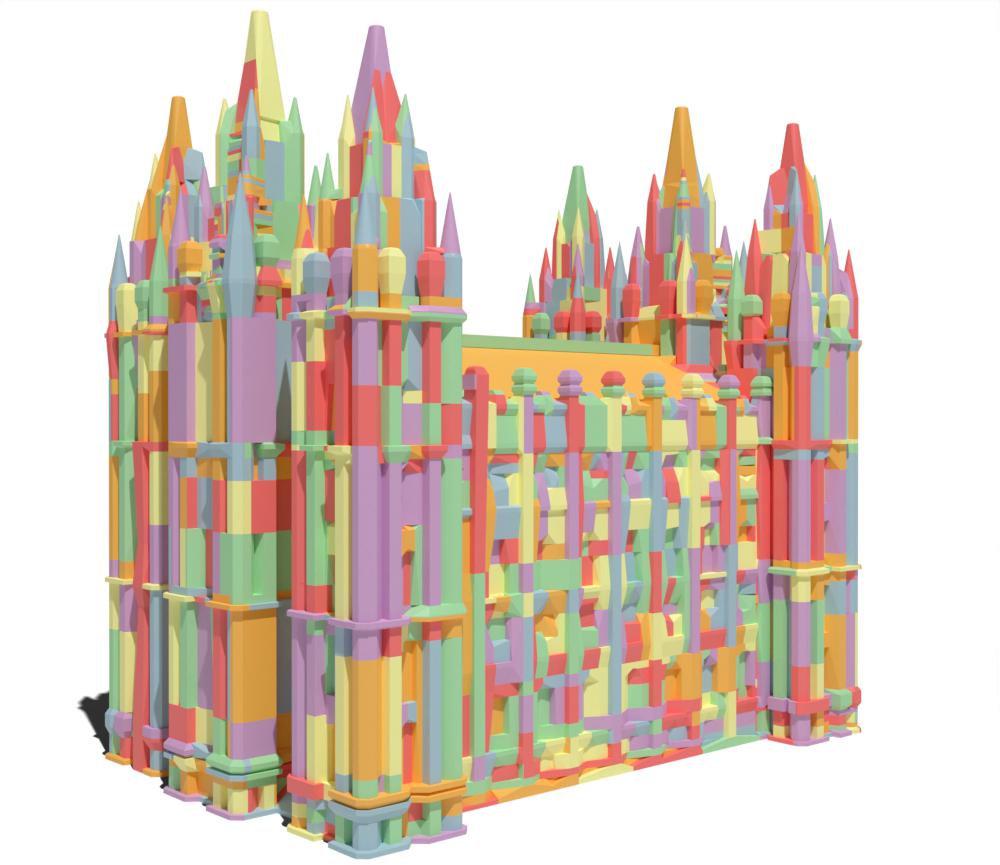}
\put(600,0){\color{yellow}\linethickness{0.35mm}%
\frame{\includegraphics[scale=.045,templeclose]{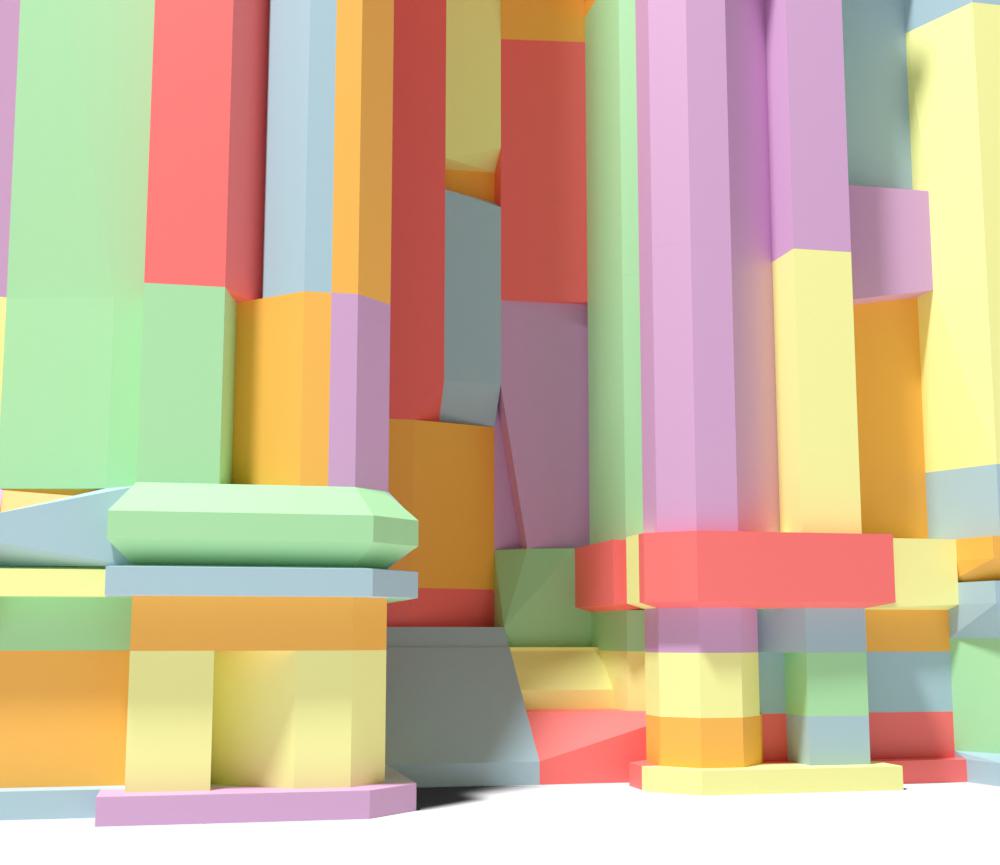}}}
\end{overpic}
&
\begin{overpic}[width=\temple,temple]
{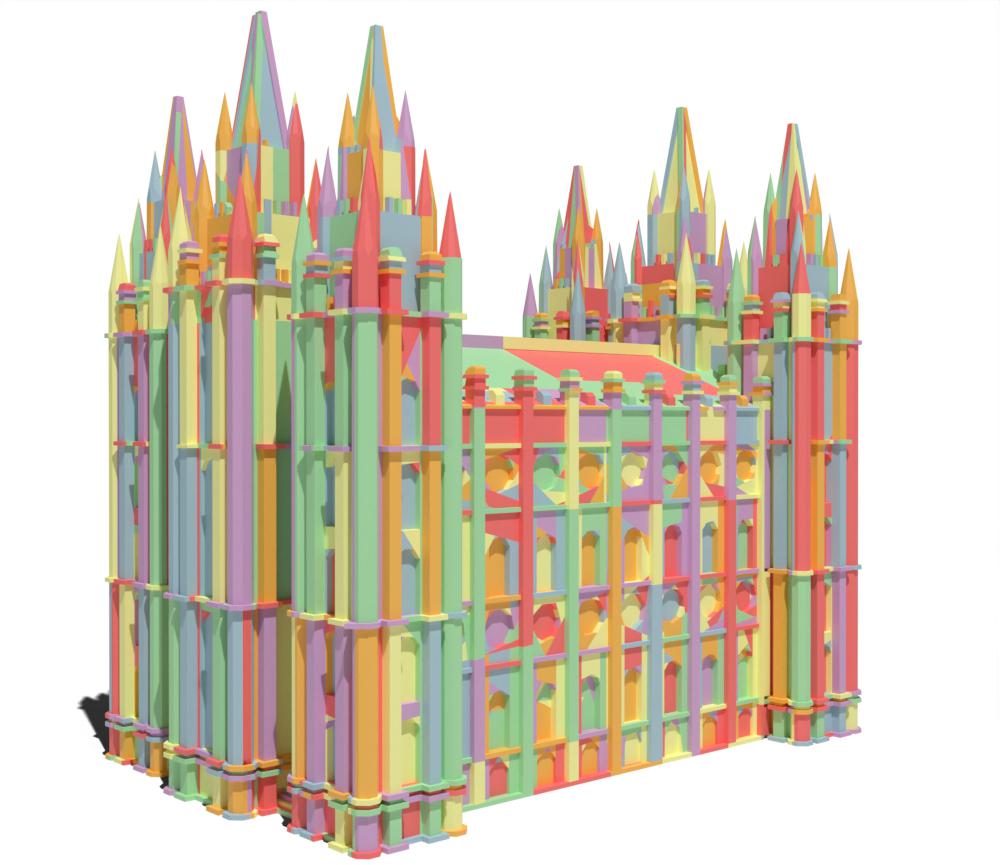}
\put(600,0){\color{yellow}\linethickness{0.35mm}%
\frame{\includegraphics[scale=.045,templeclose]{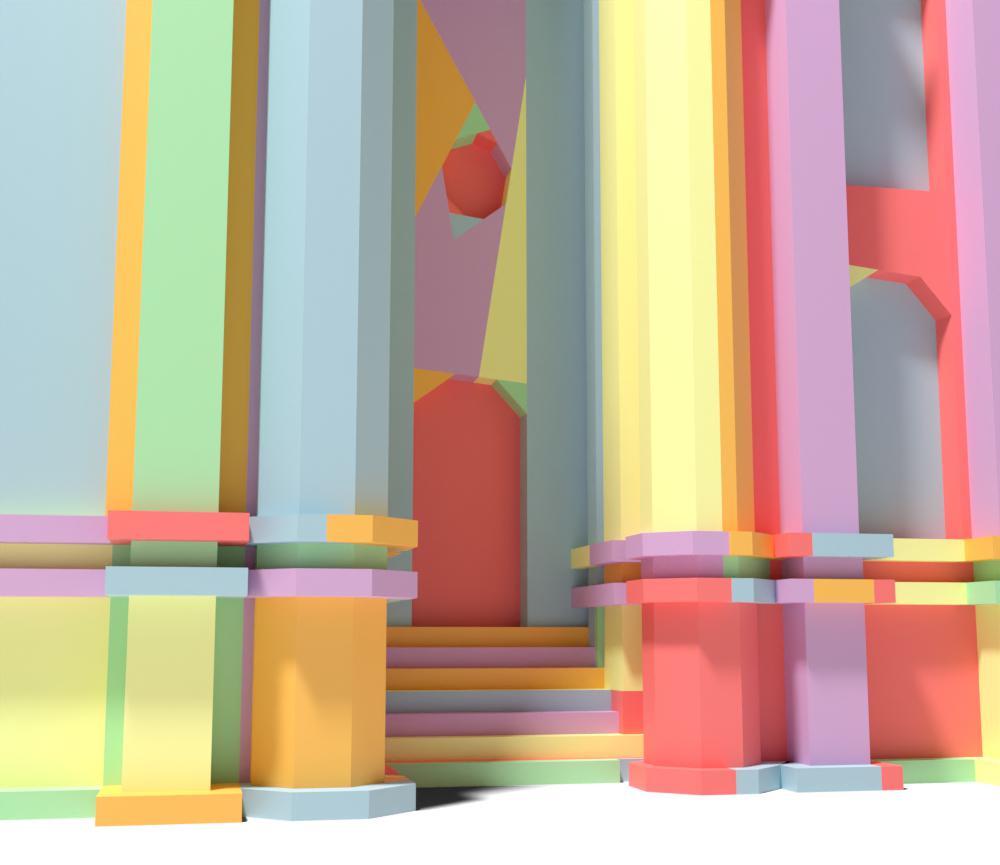}}}
\end{overpic}
\\
\myfontsize Input & \myfontsize CoACD~\cite{wei2022coacd} & \myfontsize Ours
\end{tabular}

%% file: figures/bspnet/combined_eccv/combined.tex
\begin{figure}
    
\begin{tabular}{@{}cc@{}}

\begin{subfigure}{0.28\linewidth}
    \input{figures/bspnet/combined_eccv/table.tex}
    \caption{Quantitative}
\end{subfigure}
&
\begin{subfigure}{0.72\linewidth}
    \input{figures/bspnet/combined_eccv/figure.tex}
    \caption{Qualitative}
\end{subfigure}

\end{tabular}
\caption{\textbf{Comparison on Volume Decomposition with Intersecting Convexes.} 
BSP-Net produces highly overlapping convexes (see the z-fighting) and exhibits generalisation issues (see right armrest of the sofa). Our approach produces weakly overlapping convexes that more accurately capture the fine components of the models. Note, how each pillow is described by a single convex.}
\label{fig:bspnet}
\end{figure}

%% file: figures/bspnet/combined_eccv/table.tex
\resizebox{\linewidth}{!}{%
\begin{tabular}{@{}c@{}}
    \setlength{\tabcolsep}{7pt}
\begin{tabular}{@{}lrr}
    \toprule
    & \multicolumn{2}{c}{\emph{Complexity}}  \\
    \midrule
    & $|C_V|$ & $|F_V|$  \\
     &  &  \\
    \midrule
    \textbf{BSP-Net~\cite{chen2020bspnet}}  & 60.9 & 1646   \\
    \textbf{Ours}   & \bf 41.7  & \bf 773   \\
    \bottomrule
\end{tabular}
\\[1.5cm]
\setlength{\tabcolsep}{9pt}
\begin{tabular}{rr@{}}
    \toprule
    \multicolumn{2}{c}{\emph{Accuracy}}  \\
    \midrule
    \chamfersq~$\downarrow$ & \normal~$\uparrow$ \\
    ($\times 10^{3}$) &   \\
    \midrule
    0.771 & 0.817   \\
    \bf 0.385  & \bf 0.919  \\
    \bottomrule
\end{tabular}
\end{tabular}}

%% file: figures/bspnet/combined_eccv/figure.tex
\definetrim{mytrim1}{0 0 0 0}
\definetrim{mytrim2}{100 10 100 10}
\definetrim{mytrim3}{100 50 100 50}
\definetrim{mytrim4}{0 0 0 0}
\newcommand{\mywidth}{0.33\columnwidth}
\begin{tabular}{@{}c@{}c@{}c@{}}
\includegraphics[width=\mywidth,mytrim1]{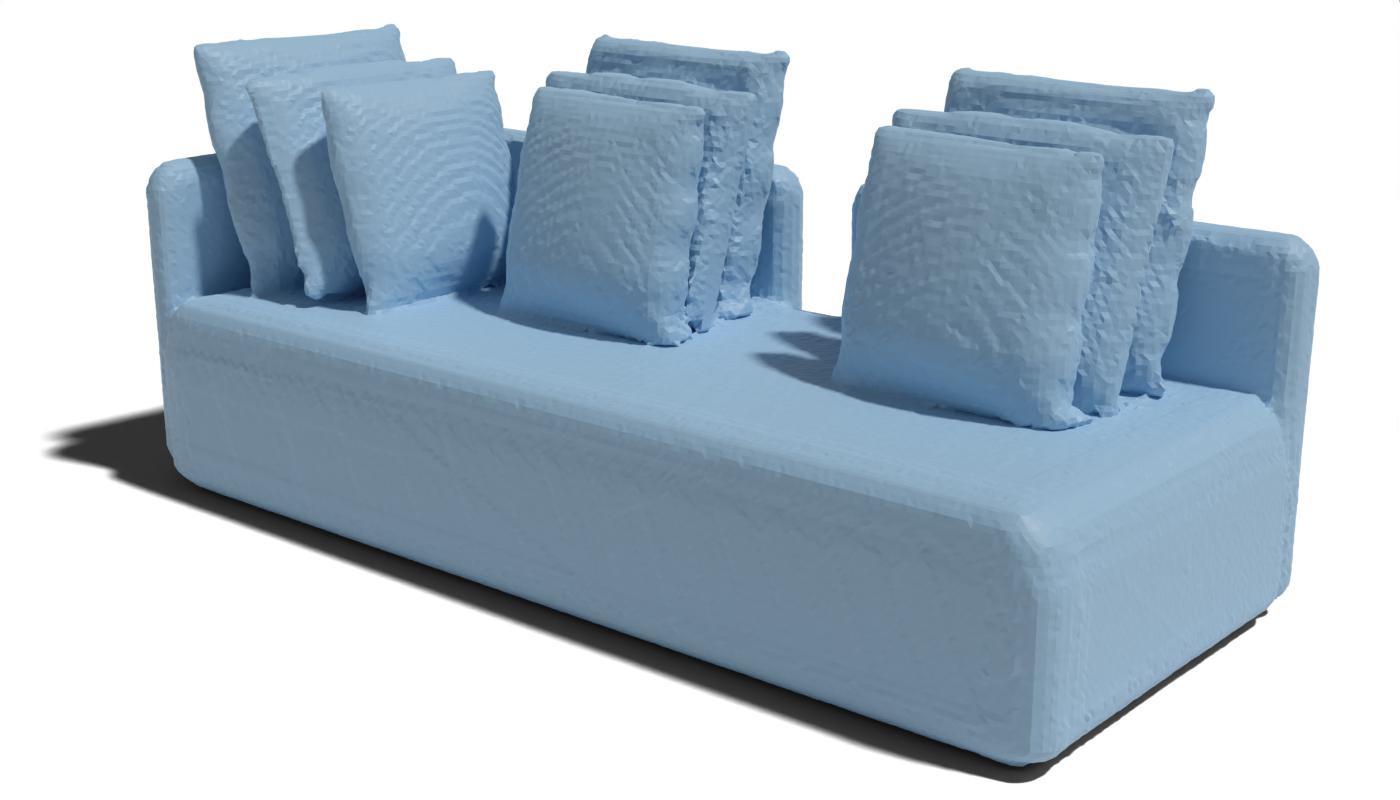}
&
\includegraphics[width=\mywidth,mytrim1]{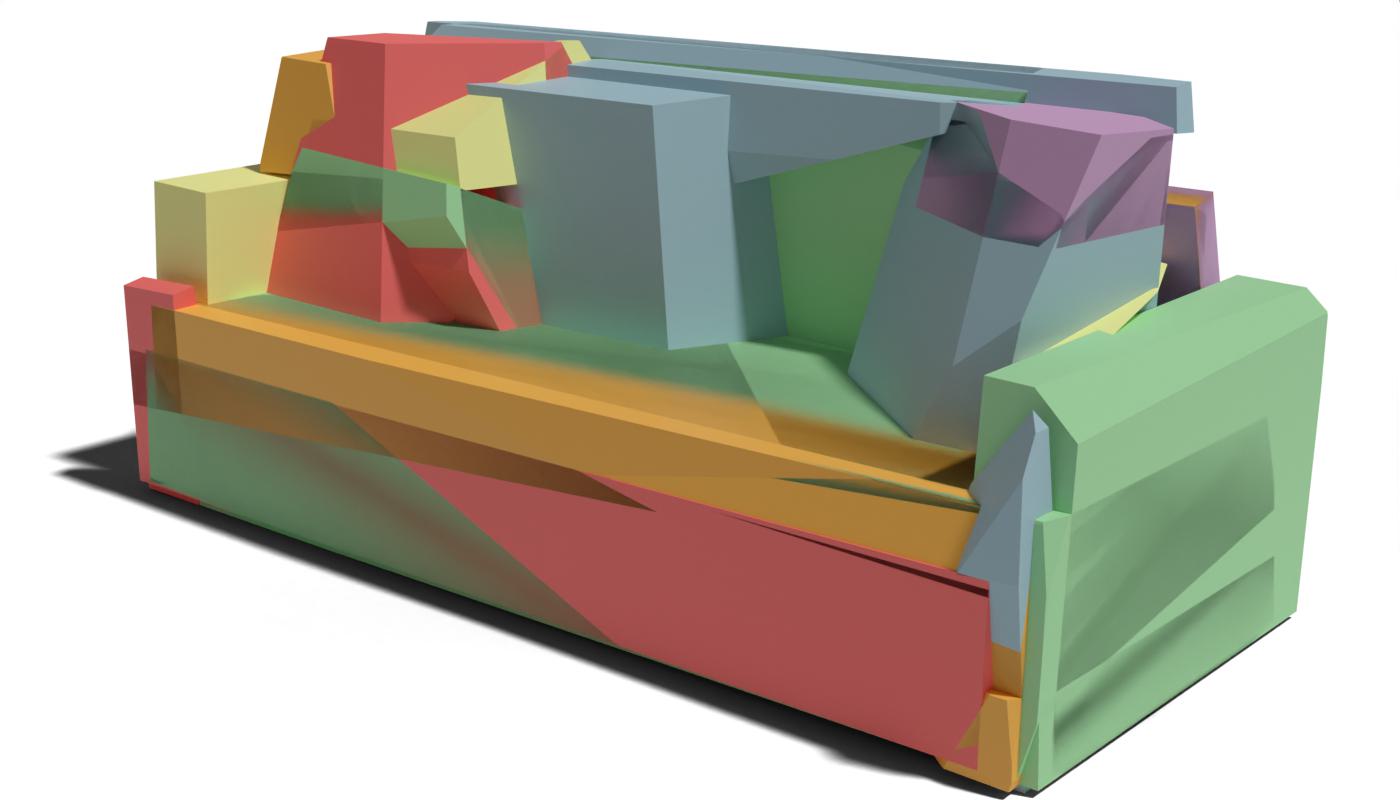}
&
\includegraphics[width=\mywidth,mytrim1]{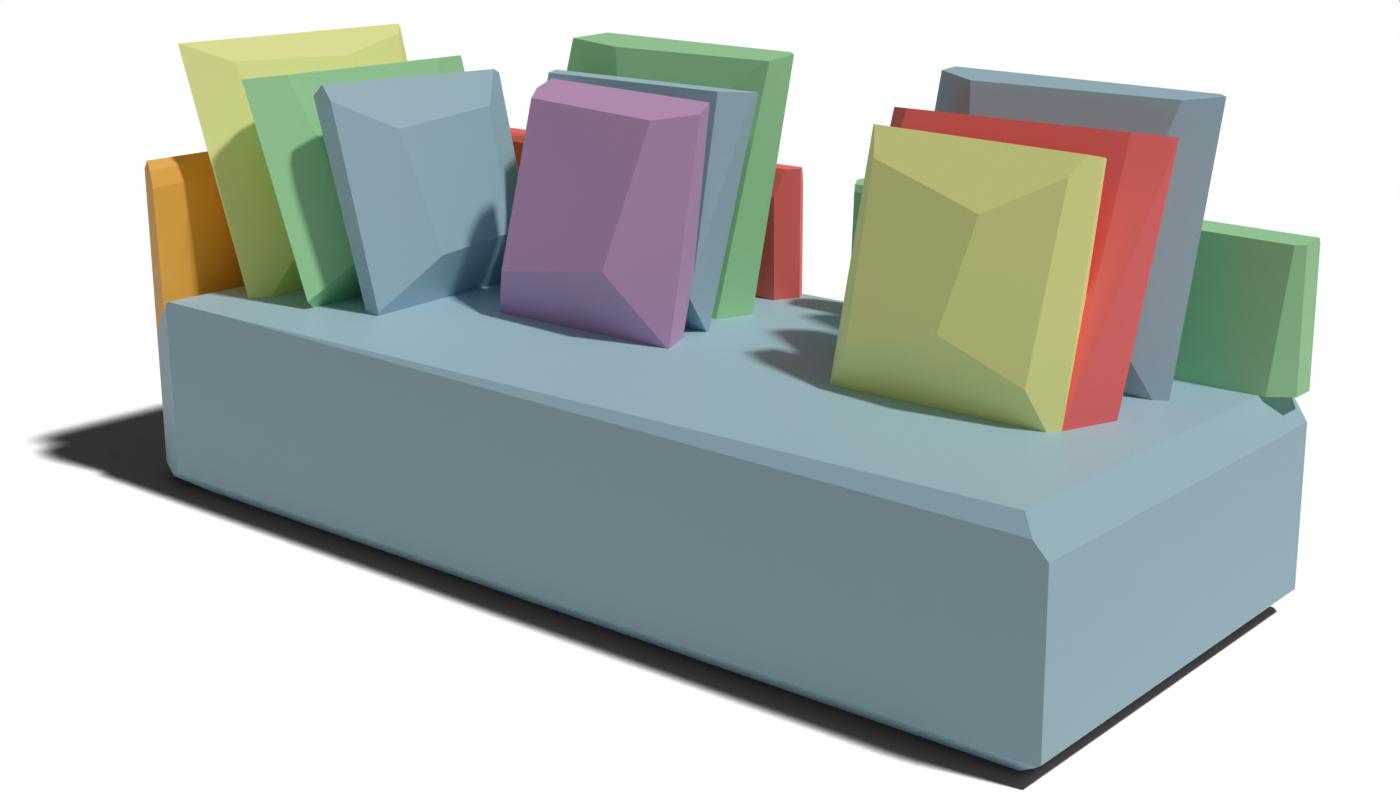}
\\
\includegraphics[width=\mywidth,mytrim1]{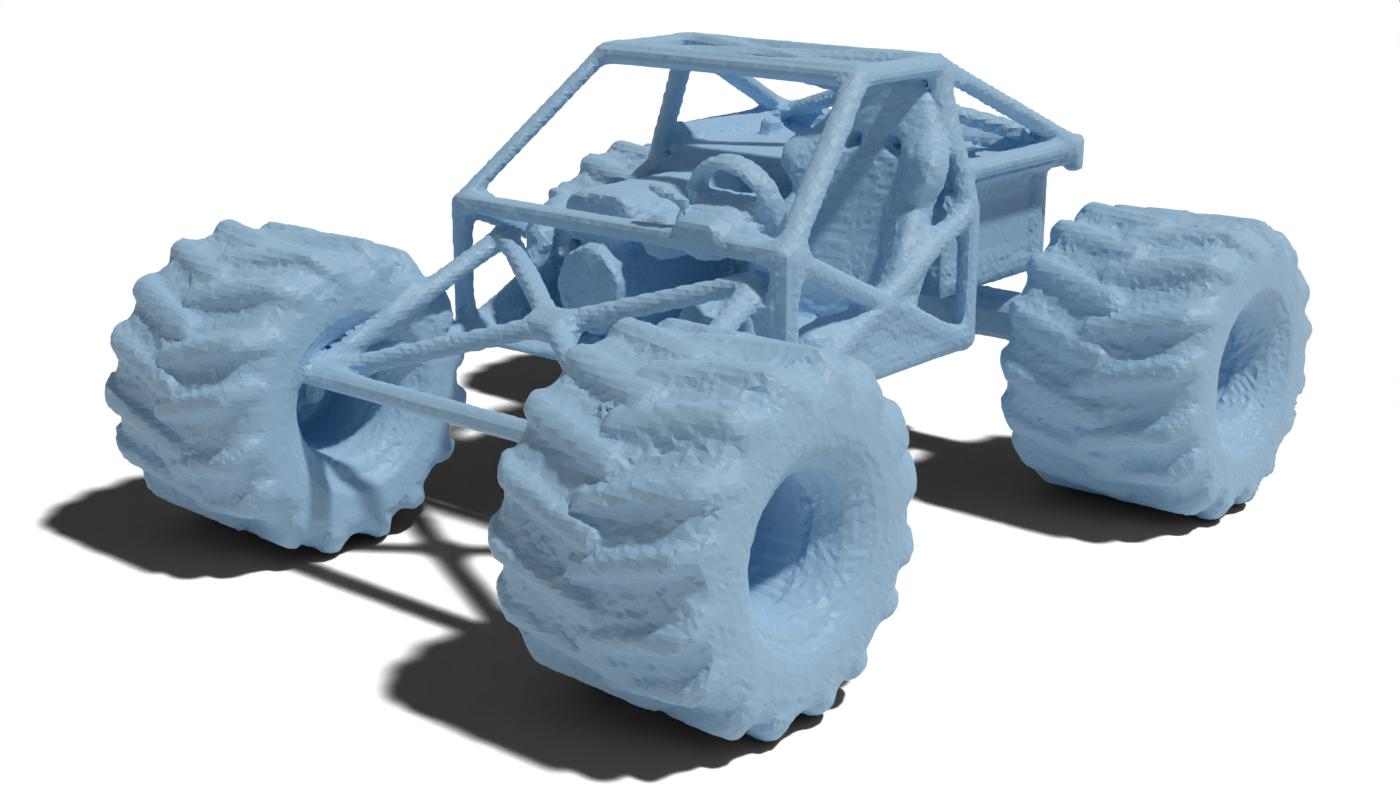}
&
\includegraphics[width=\mywidth,mytrim1]{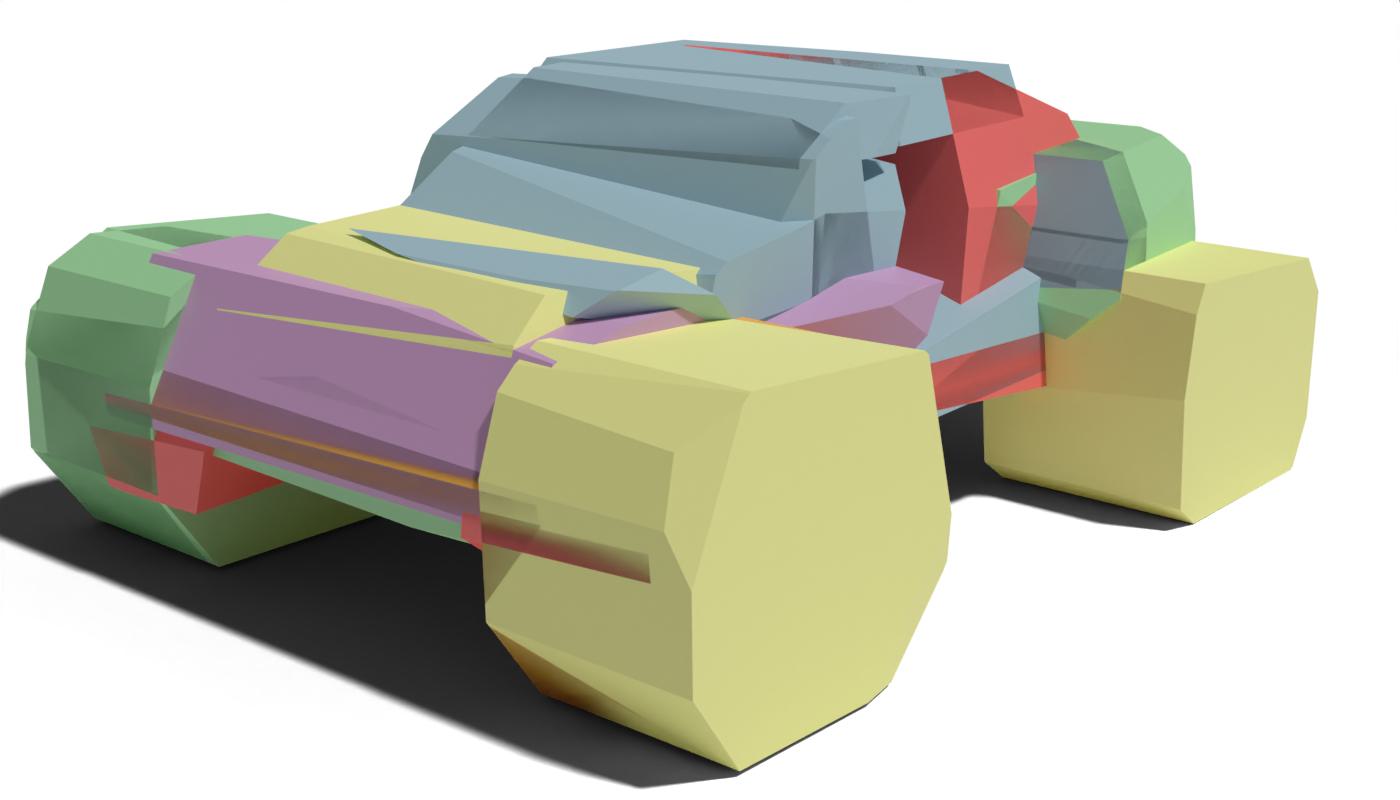}
&
\includegraphics[width=\mywidth,mytrim1]{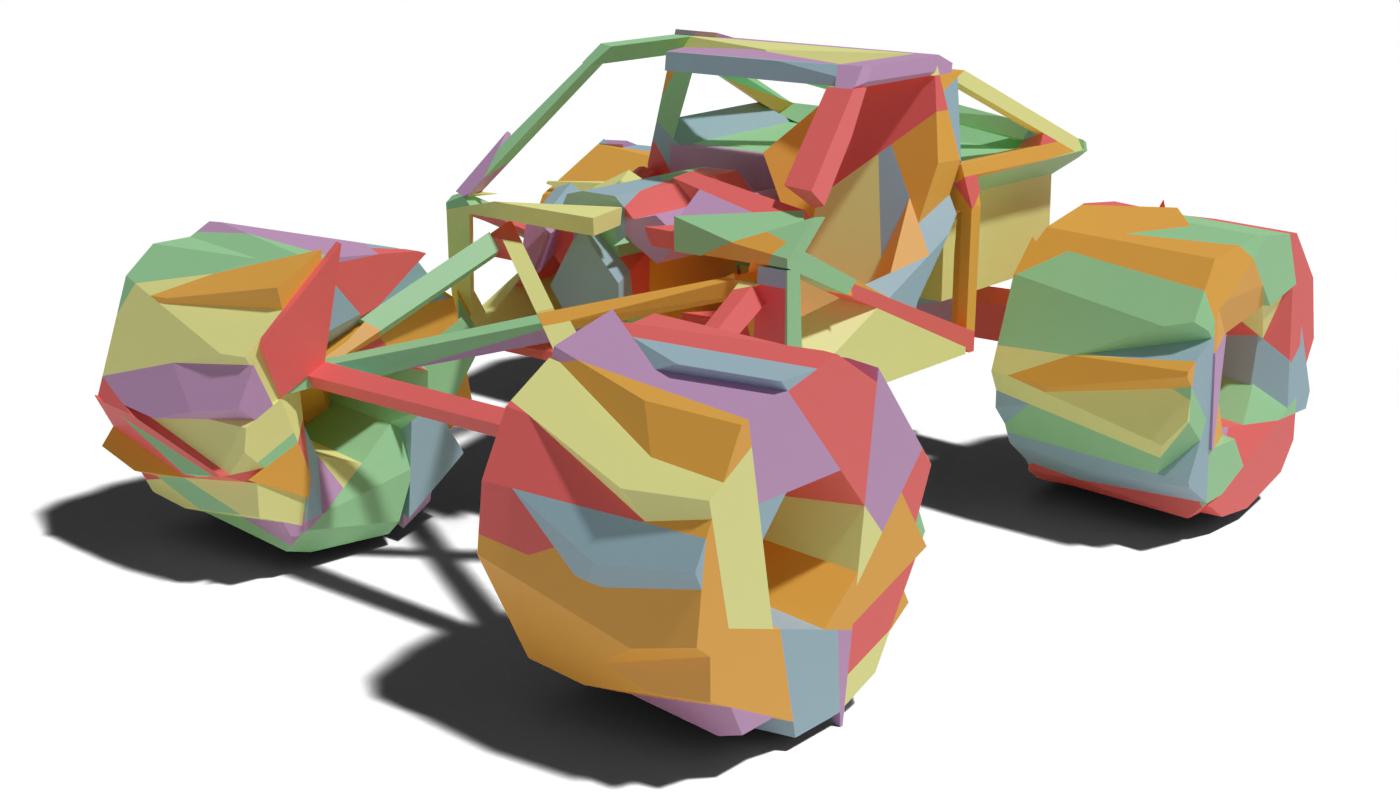}
\\
Input &BSP-Net~\cite{chen2020bspnet} & Ours
\end{tabular}

%% file: sections/5_conclusion.tex
\section{Conclusion}
We propose a scalable method for converting point clouds into concise plane arrangements. We construct our plane arrangement using (i) a specific ordering of splitting operations, (ii) input points for fast intersection queries and (iii) BSP-tree properties for simplifying the arrangement.
These steps significantly reduce the computational complexity of the construction algorithm while producing concise and meaningful arrangements. We also introduce methods to extract lightweight polygonal surface and volume meshes composed of a low number of convexes from plane arrangements.
 We demonstrated the efficiency, scalability and competitiveness of our algorithm on different low-poly mesh reconstruction problems against the best methods in the field.
In future work, we will investigate how to jointly detect and arrange planes from 3D data measurements.